%% file: main.tex
\documentclass[10pt,twocolumn,letterpaper]{article}

\usepackage[pagenumbers]{cvpr}

\input{preamble}

\definecolor{cvprblue}{rgb}{0.21,0.49,0.74}
\usepackage[pagebackref,breaklinks,colorlinks,allcolors=cvprblue]{hyperref}
\usepackage{multirow}
\usepackage{colortbl}
\usepackage{makecell}
\usepackage{algorithm}
\usepackage{algpseudocode}
\usepackage{mathtools}
\usepackage{tabularx}
\usepackage{subcaption}

\definecolor{nicered}{RGB}{174,66,38}
\definecolor{nicegreen}{RGB}{50, 140, 80}
\definecolor{cvprblue}{rgb}{0.21,0.49,0.74}

\def\confName{CVPR}
\def\confYear{2026}

\title{Language-Free Generative Editing from One Visual Example}

\author{Omar Elezabi \quad Eduard Zamfir \quad Zongwei Wu\thanks{Corresponding Author} \quad Radu Timofte\\
{\small{Computer Vision Lab, CAIDAS \& IFI, University of Würzburg}}\\
}

\begin{document}
\maketitle
\input{sec/0_abstract}    
\input{sec/1_intro}
\input{sec/2_rw}

\input{sec/3_method}
\input{sec/4_experiments}
\input{sec/4.2_Ablation}
\input{sec/5_conclusion}

\input{sec/supp}

{
    \small
    \bibliographystyle{ieeenat_fullname}
    \bibliography{main}
}

\end{document}

%% file: preamble.tex
\newcommand*{\myparagraphnospace}[1]{\noindent\textbf{#1}\hspace{2mm}}

\definecolor{cvprblue}{rgb}{0.21,0.49,0.74}
\definecolor{pltblue}{RGB}{174, 199, 232}
\definecolor{pltorange}{RGB}{255, 229, 204}
\definecolor{pltgreen}{RGB}{204, 229, 204}
\definecolor{pltred}{RGB}{229, 204, 204}
\definecolor{pltpurple}{RGB}{239, 218, 230}

\definecolor{tabblue}{HTML}{1f77b4}
\definecolor{taborange}{HTML}{ff7f0e}
\definecolor{tabgreen}{HTML}{2ca02c}
\definecolor{tabred}{HTML}{d62728}
\definecolor{tabpurple}{HTML}{9467bd}

\definecolor{cblue}{RGB}{173, 201, 233}
\definecolor{clblue}{RGB}{222, 234, 246}
\definecolor{corange}{RGB}{255, 152, 67}
\definecolor{lorgange}{RGB}{255, 221, 149}



%% file: sec/0_abstract.tex
\vspace{-3mm}
\begin{abstract}
Text-guided diffusion models have advanced image editing by enabling intuitive control through language.
However, despite their strong capabilities, we surprisingly find that SOTA methods struggle with simple, everyday transformations such as rain or blur. 
We attribute this limitation to weak and inconsistent textual supervision during training, which leads to poor alignment between language and vision. 
Existing solutions often rely on extra finetuning or stronger text conditioning, but suffer from high data and computational requirements. 
We argue that diffusion-based editing capabilities aren't lost but merely hidden from text.
The door to cost-efficient visual editing remains open, and the key lies in a vision-centric paradigm that perceives and reasons about visual change as humans do, beyond words.
Inspired by this, we introduce \textbf{Visual Diffusion Conditioning} (VDC), a training-free framework that learns conditioning signals directly from visual examples for precise, language-free image editing.
Given a paired example—one image with and one without the target effect—VDC derives a visual condition that captures the transformation and steers generation through a novel condition-steering mechanism.
An accompanying inversion-correction step mitigates reconstruction errors during DDIM inversion, preserving fine detail and realism.
Across diverse tasks, VDC outperforms both training-free and fully fine-tuned text-based editing methods.
The code and models are open-sourced at \href{https://omaralezaby.github.io/vdc/}{\texttt{omaralezaby.github.io/vdc/}}
\end{abstract}

%% file: sec/1_intro.tex
\section{Introduction}

Diffusion models have revolutionized visual synthesis, powering the current state-of-the-art in image editing~\cite{ho2020denoising,song2020denoising,dhariwal2021diffusion}. 
Notably, text-guided diffusion models enable intuitive manipulation through natural language prompts~\cite{rombach2022high,xie2024sana,dai2023emu}, offering strong spatial and semantic control.

\begin{figure}[t]
    \centering
    \vspace{-3mm}
    \includegraphics[width=\columnwidth]{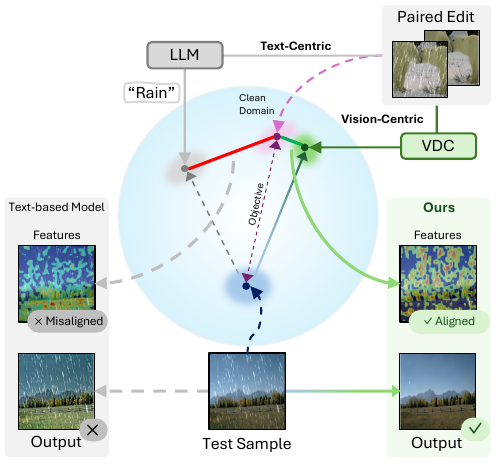}
    \vspace{-5mm}
    \caption{
    \textit{Text–image misalignment in diffusion latent space.}
Text-guided generative models rely on language, which often fails to capture appearance-level transformations, \textit{e.g.} rain, leading to semantic but visually \textit{\textcolor{nicered}{misaligned}} directions.
Our method, \textbf{Visual Diffusion Conditioning} (VDC), instead learns a vision-centric conditioning signal directly from paired visual examples, uncovering the \textit{\textcolor{nicegreen}{correct}} transformation direction within the latent space.
By steering the diffusion process along this aligned path, VDC achieves faithful and realistic edits, bridging the gap between text semantics and visual representations.}
    \vspace{-6mm}
    \label{fig:intro:teaser}
\end{figure}

Despite their impressive flexibility, we find that current text-guided diffusion models often struggle with simple visual transformations such as rain, haze or blur.
As we illustrate in \cref{fig:intro:teaser}, their internal representations fail to match the semantics of these textual descriptions.
We link this behavior to weak and inconsistent supervision: diffusion models rely on image–caption pairs and thus learn only the concepts explicitly described in the training data.
Consequently, visual phenomena that are rarely or ambiguously captioned exhibit poor alignment between text prompts and their associated visual features.

A natural solution might be to fine-tune the model for these missing concepts~\cite{brooks2023instructpix2pix,xiao2025omnigen,zhang2025context,li2025superedit}.
However, retraining large diffusion models is computationally intensive and data-hungry, rendering it impractical for most editing scenarios.
Importantly, diffusion models already encode rich, structured visual representations that extend beyond their textual supervision~\cite{kwon2023diffusion}.

The limitation arises from weak language-vision alignment, which obscures access to the full visual manifold, as shown in \cref{fig:intro:feature_vis}.
We propose to bridge this gap through a \textit{vision-centric} perspective on editing.
Rather than relying on language to approximate visual intent, our approach treats manipulation as a process grounded in perceptual change.
Visual examples—unlike text—can unambiguously express such changes: a pair of images naturally encodes degradations, or stylistic variations that are difficult to capture verbally.
By extracting conditioning signals directly from visual examples, we can translate observable differences into latent-space directions that operate on the model’s existing visual representations, \textit{c.f.} \cref{fig:intro:feature_vis}.
This motivates an editing framework driven by \textit{visual exemplars} instead of language.

Building on this, we introduce \textbf{Visual Diffusion Conditioning} (VDC), a training-free diffusion editing framework that learns visual conditioning signals from example image pairs.
Instead of text prompts, VDC derives a compact representation that encodes the transformation between two visual domains (\textit{e.g.}, clean $\leftrightarrow$ degraded).
Once extracted, this visual condition can be transferred to unseen images, enabling consistent and controllable edits.
Prior training-free methods~\cite{mokady2023null,miyake2025negative,hertz2022prompt,kim2025reflex,sun2025attentive,tumanyan2023plug,cao2023masactrl,liu2023more,meng2021sdedit,wallace2023edict,ju2023direct,wang2024taming,xu2023inversion} typically operate by inverting the diffusion process and modifying latent trajectories through textual guidance.
While effective for semantic manipulation, they remain limited by language–vision misalignment and struggle to express fine-grained, appearance-level changes.
Besides, current exemplar-driven approaches~\cite{vsubrtova2023diffusion,nguyen2023visual,gu2024analogist,wang2025editclip,kim2025difference} partially address this issue by defining edits from image pairs, but most rely on pretrained vision–language models~\cite{radford2021learning} or additional finetuning~\cite{wang2025editclip}, which reduces generality and increases computational cost.

In contrast, our VDC framework introduces pure visual conditioning, leveraging the pretrained latent structure.

Our framework builds on two core components:
(i) a \textit{condition steering mechanism} that modulates the sampling process via posterior score guidance~\cite{song2020score}, enabling precise and stable edits without retraining; and
(ii) an \textit{inversion correction step} that compensates for error accumulation in DDIM inversion~\cite{song2020denoising,dhariwal2021diffusion}, preserving perceptual quality.
In summary, our main contributions are:
\begin{itemize}
    \item A diffusion editing framework, termed \textit{\textbf{V}isual \textbf{D}iffusion \textbf{C}onditioning}, that learns directly from visual examples.
    \item A stable, lightweight neural embedding that captures edit semantics from a single example pair, enabling training-free yet generalizable editing.
    \item A sampling and inversion strategy that achieves precise editing while preserving perceptual fidelity.
\end{itemize}

%% file: sec/2_rw.tex
\section{Related Works}

\begin{figure}[t]
    \centering
    \includegraphics[trim={.5cm 0 0 0},width=0.95\columnwidth]{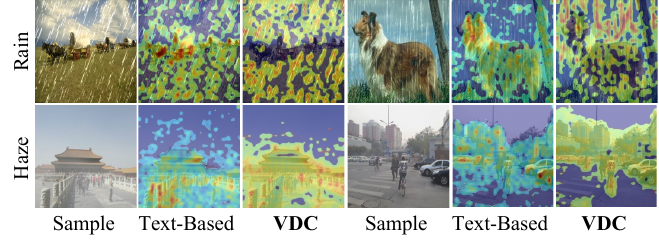}
    \vspace{-4mm}
    \caption{\textit{Language-Vision misalignment.} 
    The internal representations of LDM~\cite{rombach2022high} fail to accurately capture the semantics of degradations such as “rain” or “haze”.
    Attention maps under text-based conditioning remain object-centric and do not correspond to degradation-specific visual attributes.
    Our VDC framework realigns attention focus toward true visual cues, recovering meaningful features that correspond to rain streaks and hazy regions.
    }
    \vspace{-5mm}
    \label{fig:intro:feature_vis}
\end{figure}

\begin{figure*}[t]
    \centering
    \begin{subfigure}{\columnwidth}
        \centering
        \includegraphics[width=\columnwidth]{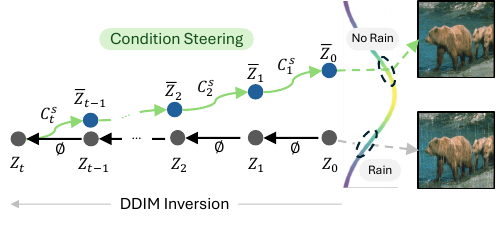}
        \caption{\textit{DDIM inversion with Condition Steering.}}
        \label{fig:meth:subfig_a}
    \end{subfigure}%
    \hfill
    \begin{subfigure}{\columnwidth}
        \centering
        \includegraphics[width=\columnwidth,]{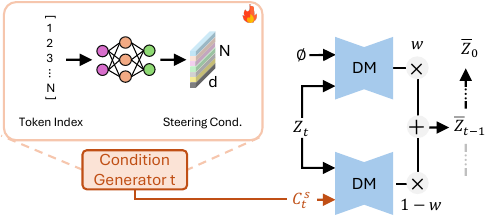}
        \caption{\textit{Steering Condition Generation.}}
        \label{fig:meth:subfig_b}
    \end{subfigure}
    \vspace{-3mm}
    \caption{\textit{Proposed VDC framework.} 
    (a) Given a real image, we first invert it through DDIM and apply the learned steering condition $C_t^s$ to guide sampling toward the desired visual feature (e.g., removing rain) while preserving content and quality.
    (b) A lightweight \textit{Condition Generator} produces per-step steering embeddings from token indices, representing the target visual feature.
    These conditions modulate the diffusion outputs through weighted score blending, enabling training-free visual editing without textual prompts.}
    \label{fig:meth:main_diagram}
    \vspace{-5mm}
\end{figure*}

\myparagraphnospace{Text-based Image Editing.}
Instruction-based image editing methods \cite{brooks2023instructpix2pix} were proposed to modify an input image according to text instructions. These approaches typically employ generative models to synthesize large-scale instruction-based editing datasets, which are then used to fine-tune diffusion models for conditional image editing. Subsequent works refined this paradigm by curating higher-quality datasets \cite{zhang2023magicbrush} and leveraging improved architectures and generative backbones \cite{xiao2025omnigen,zhang2025context,li2025superedit,labs2025flux,sheynin2024emu}.
To reduce the dependence on large-scale instruction data and computationally expensive training, train-free methods \cite{liu2023more,meng2021sdedit,kawar2023imagic,hertz2022prompt,wallace2023edict,sun2025attentive,kim2025reflex,tumanyan2023plug,mokady2023null,ju2023direct,cao2023masactrl,miyake2025negative,wang2024taming,xu2023inversion} were introduced. These methods exploit the intrinsic generative and semantic capabilities of pretrained text-to-image (T2I) diffusion models to perform edits without retraining. They typically invert the diffusion process \cite{song2020denoising,dhariwal2021diffusion,ju2023direct,wang2024taming,xu2023inversion} to recover the latent noise representation of an input image, and then modify conditioning components such as the textual prompt \cite{kawar2023imagic,mokady2023null,miyake2025negative}, self-attention maps \cite{sun2025attentive,tumanyan2023plug,cao2023masactrl}, or cross-attention modules \cite{hertz2022prompt,kim2025reflex} to realize desired edits. Despite their flexibility, purely text-based methods often struggle to capture fine-grained or compositional edits that go beyond what can be easily expressed in language.

\vspace{1mm}
\myparagraphnospace{Exemplar-based Editing.}
A core limitation of text-based editing lies in its reliance on natural language, which is often ambiguous and insufficient for describing complex, localized, or stylistic edits. To address this, visual exemplar-based editing methods incorporate visual examples to define edits more precisely \cite{vsubrtova2023diffusion,nguyen2023visual,gu2024analogist,wang2025editclip,kim2025difference}. These approaches learn from pairs of “before” and “after” example images to infer a transformation that can be applied to new inputs. Typically, they employ textual or joint vision–language representations to model the relationship between the visual example pair and the input image. However, even these methods depend on text-aligned latent spaces, inheriting the limitations of T2I diffusion models, such as the imperfect alignment between textual embeddings and visual features.
Although some works attempt to fine-tune diffusion models directly for the visual instruction setting, they still rely on VLMs \cite{radford2021learning} to extract edit semantics \cite{wang2025editclip,luo2023controlling}. This dependence often leads to the loss of global context or fine visual details, constraining edit fidelity and controllability.

\vspace{1mm}
\myparagraphnospace{Diffusion for Inverse Problems.}
Diffusion models have also been successfully applied to inverse problems \cite{chung2022diffusion,wang2022zero,zhu2023denoising,fei2023generative,chung2022improving} due to their powerful ability to model complex data distributions. By reformulating image restoration as a guided sampling task, diffusion models can recover clean images that correspond to a given degraded observation—achieving zero-shot restoration without additional training. Initially introduced for image-space diffusion models, these approaches were later extended to latent diffusion models to better exploit their semantic priors and efficiency \cite{rout2023solving,kim2023regularization,song2023solving,rout2024beyond,xiao2024dreamclean,zhang2025improving}. Nevertheless, these methods typically assume known degradation operators (\textit{e.g.}, blur kernels, noise levels), which limits their generalization to complex, spatially varying degradations such as haze, rain, or reflection removal.

%% file: sec/3_method.tex
\section{Methodology}

\begin{algorithm}
\caption{Steering Condition Generator Optimization}\label{alg:cap}
\noindent\textbf{Input:} $R_B$ Visual Example Before Editing, $R_A$ Visual Example After Editing, $Itrs$ Number of optimization iterations, $p\sim[T,0)$ Diffusion step for resampling start, $N$ Number of tokens in the condition, $\phi$ Null Condition, $E$ and $D$ Encoder and Decoder respectively. \\
\textbf{Output:} Optimized Steering Condition $C^s$\\
$Z^B, Z^A = E(R_B), E(R_A)$

\Comment{partial inversion to step p}
 
$Z^B_p = \text{DDIM}_{\text{Inversion}}(Z^B, t=(1,...p), \phi)$

\begin{algorithmic}
\For{$i=1,...Itrs$}
\For{$t=p,...1$}

\Comment{generate step condition}

$C^s_t$ = $\text{MLP}_t(1,...N)$ 

$\epsilon_{init} = \epsilon_{\theta}(Z^B_t, t, \phi)$ \Comment{adjust sampling}

$\epsilon_{\text{steering}} = \epsilon_{\theta}(Z^B_t, t, C^s_t)$

$\hat\epsilon = (1-w) * \epsilon_{\text{steering}} + w * \epsilon_{\text{init}}$

$Z^B_{t-1} = \text{DDIM}_{\text{Step}}(Z^B_t, t, \hat\epsilon)$
\EndFor

\Comment{Optimize Condition Generator}

$ \mathcal{L} = || Z^B_{0} - Z^A ||_2^2 + || D(Z^B_{0}) - R_A ||_2^2$ 

$ \text{MLP}_{1,...t} = \text{MLP}_{1,...t} + \text{AdamGrad}(\mathcal{L})$ 
\EndFor
\end{algorithmic}

\textbf{return} $C^s \gets \text{MLP}_{1,...t}(0,...N)$
\end{algorithm}

\begin{algorithm}

\caption{DDIM Inversion Correction}\label{alg:ddim}
\noindent\textbf{Input:} $Z_0$ latent to be inverted , $I$ number of iterations, $p\sim[T,0)$ Diffusion resampling start, $\phi$ Null Condition \\
\textbf{Output:} Corrected Noised Latent $z^*_p$
 
$\bar{Z}_p = \text{DDIM}_{\text{Inversion}}(Z_0, t=(1, ...p), \phi)$

\begin{algorithmic}

\For{$i=1,...I$}

$\hat{Z}_0 = \text{DDIM}_{\text{Forward}}(\bar{Z}_p, t=(p, ...1), \phi)$

$\mathcal{L} = || \hat{Z}_0 - Z_0 ||_2^2$

$\bar{Z}_p =  \bar{Z}_p - \text{AdamGrad}(\mathcal{L})$

\EndFor

\end{algorithmic}

\textbf{return} $\bar{Z}_p$

\end{algorithm}

\paragraph{Diffusion Preliminaries.}
Diffusion models generate data by iteratively denoising a latent variable $z_t$ sampled from a Gaussian prior.
At each timestep $t$, a noise prediction network $\epsilon_{\theta}(z_t, t, C)$ estimates the denoised sample conditioned on $C$, which can be a text embedding or other guidance signal.

Inversion methods such as DDIM~\cite{song2020denoising} allow reconstructing a latent trajectory from a real image, enabling editing in latent space.
As shown in \cref{fig:meth:subfig_a}, our framework builds on these foundations by replacing textual conditioning with a learned \textit{visual condition}, used to steer the generative process toward appearance-level transformations.

\subsection{Editing by Visual Conditioning}

VDC builds on the observation that diffusion models implicitly recognize visual features even when these features lack corresponding textual representations.
Although text prompts fail to access such features, they can be revealed by shifting from language-based to purely visual conditioning.
We achieve this by identifying a condition that captures a specific transformation through visual examples.
Given an image pair before and after editing, $(R_B, R_A)$, we derive a visual condition $C^s$ that encodes the transformation within the model’s learned data distribution, as shown in \cref{fig:meth:subfig_b}.
By inverting a real image and applying this condition during the generative process, we steer the model to reproduce the desired edit.
This enables representation and manipulation of visual features without textual prompts, unlocking the full expressive capacity of the diffusion latent space.

\subsection{Condition Steering}

To completely detach from the textual space, we consider an unconditional generative process and manipulate the image by steering the sampling trajectory using a condition that represents the visual feature to be edited or removed (\textit{e.g.}, rain, fog, or noise).
Given a condition representing a visual feature $C^s$, we steer the generative process according to the posterior score function~\cite{song2020score} of the unconditional model:
\vspace{-2mm}
\begin{equation}
    \nabla_x \log p(x|C^s) = \nabla_x \log p(x) +  \nabla_x \log p(C^s|x)
    \vspace{-2mm}
\end{equation}

For tasks such as deraining or dehazing, where $C^s$ denotes the feature to be removed, the goal is to steer sampling away from the high-density region of that feature.
This can be expressed as the posterior score function for $-C^s$:
\vspace{-2mm}
\begin{equation}
    \nabla_x \log p(x|-C^s) = \nabla_x \log p(x) -  s *\nabla_x \log p(C^s|x)
    \vspace{-2mm}
\end{equation}
Here, $s$ is a hyperparameter controlling the steering intensity, and by Bayes’ rule, $p(C^s|x) \sim p(x|C^s)/p(x)$.
Expanding this relation gives:
\vspace{-2mm}
\begin{flalign}
    \nonumber &\nabla_x \log p(x|-C^s) =\\
    &\nabla_x \log (x) -  s *(\nabla_x \log p(x|C^s) - \nabla_x \log p(x) )
\end{flalign}
\vspace{-5mm}

\noindent Adapting this to the noise prediction model in LDM, where $\nabla_x \log (x) \sim \epsilon_{\theta}(z_{t}, t, \phi)$ and $\log p(x|C^s) \sim \epsilon_{\theta}(z_{t}, t, C_t)$, we can rewrite the formulation as:
\vspace{-2mm}
\begin{flalign}
\nonumber
    \epsilon_{\theta}(z_{t}, -C^s) &= \epsilon_{\theta}(z_{t}, \phi) - s *(\epsilon_{\theta}(z_{t}, C^s) - \epsilon_{\theta}(z_{t}, \phi)) &&\\\nonumber
    &= \epsilon_{\theta}(z_{t}, C^s) + (1+s) *(\epsilon_{\theta}(z_{t}, \phi) - \epsilon_{\theta}(z_{t}, C^s))&&\\
    &= (1-w) * \epsilon_{\theta}(z_{t}, C^s) + w * \epsilon_{\theta}(z_{t}, \phi)
\end{flalign}

\noindent where $w = 1 + s$.
This formulation enables direct manipulation of the visual feature represented by $C^s$ by steering the trajectory of the unconditional generative process used to invert the real image.
Editing the image in this way avoids generative artifacts, since we update the inverted image ($out = Z(\phi) + Z(C_\theta)$) rather than generating a new image ($out = Z(C_\theta)$), analogous to a global residual connection in image-to-image networks~\cite{pang2021image}.
We visualize this process in \cref{fig:meth:subfig_a}.

\subsection{Condition Representation for Visual Features}

In diffusion models, the conditioning input is typically represented as a sequence of tokens, each corresponding to an encoded word in the textual prompt (\textit{e.g.}, Stable Diffusion~\cite{rombach2022high} accepts up to 77 tokens as input).
Optimizing textual embeddings has been used to improve diffusion inversion of real images~\cite{mokady2023null} or to personalize the generative process by learning an embedding for a specific object~\cite{ruiz2023dreambooth}.
This optimization treats text embeddings as trainable parameters and updates them according to a chosen objective function.
However, the process depends on an initial prompt embedding and is often unstable, allowing optimization of only a small number of tokens~\cite{vsubrtova2023diffusion, mokady2023null}.

To fully remove textual dependency, we generate a new embedding directly from a condition generator network.
Inspired by Implicit Neural Representations (INR)~\cite{tancik2020fourier,sitzmann2020implicit}, which encode images as continuous functions over pixel coordinates, we represent the visual edit condition as a continuous function over token indices.
Specifically, we employ a lightweight three-layer MLP and, following INR literature, apply Fourier features to the input indices to improve expressiveness~\cite{tancik2020fourier}.
This formulation provides stable optimization when learning the steering condition that represents a desired edit.
The improved stability allows optimization of all 77 tokens, enabling full access to the model’s visual condition space.
Further, since each token is generated from a continuous function conditioned on token indices, the network naturally establishes communication across tokens, producing smooth and coherent condition representations.
For finer control during editing, we optimize a separate condition generator for each diffusion step.
\vspace{-2mm}
\begin{flalign}
\label{eq:ours_single}
\nonumber &C^s_t = \text{MLP}_t(1,....N)\\
&\underset{\text{MLP}_{t}}{\text{min}}||  Z_{t-1}^* - Z_{t-1}(Z_t,t,C_t) ||_2^2
\end{flalign}

\begin{table*}[!ht]
    \centering
    \footnotesize
    \caption{\textit{Comparison to state-of-the-art image editing.} FID ($\downarrow$) and LPIPS($\downarrow$) are reported on the full RGB images. Our method sets a new state-of-the-art on average across all benchmarks. ‘-’ represents unreported results. The \textbf{best} performances are highlighted.}
    \vspace{-3mm}
    \setlength\tabcolsep{3.5pt}
    \begin{tabularx}{\textwidth}{cX*{12}{c}}
        \toprule
        \multirow{2}{*}{Type} & \multirow{2}{*}{Method} & \multicolumn{2}{c}{SR} & \multicolumn{2}{c}{DeBlur} & \multicolumn{2}{c}{DeNoise} & \multicolumn{2}{c}{DeRain} & \multicolumn{2}{c}{DeHaze} & \multicolumn{2}{c}{Colorization} \\
        {} & {} & FID $\downarrow$ & LPIPS $\downarrow$ & FID $\downarrow$ & LPIPS $\downarrow$ & FID $\downarrow$ & LPIPS $\downarrow$ & FID $\downarrow$ & LPIPS $\downarrow$ & FID $\downarrow$ & LPIPS $\downarrow$ & FID $\downarrow$ & LPIPS $\downarrow$ \\
        \midrule
        \multirow{3}{*}{\rotatebox{90}{T-Edit}} & P2P \cite{hertz2022prompt} & 126.47 & 0.6662 & 45.62 & 0.5220 & 142.95 & 0.5593 & 139.19 & 0.3122 & 44.09 & 0.2183 & 121.87 & 0.2931 \\
        {} & Null-Opt \cite{mokady2023null} & 73.48 & 0.5510 & 51.89 & 0.5258 & 160.88 & 0.6059 & 167.61 & 0.5050 & 91.76 & 0.4917 & 197.81 & 0.5881 \\
        {} & Negative-Cond \cite{miyake2025negative} & 63.22 & 0.4807 & 43.61 & 0.4528 & 96.19 & 0.4764 & 118.76 & 0.3157 & 43.20 & 0.2193 & 135.63 & 0.3407 \\
        \midrule
        \multirow{4}{*}{\rotatebox{90}{I-Edit}} & Instruct-Pix2Pix \cite{brooks2023instructpix2pix} & 92.79 & 0.5828 & 142.91 & 0.7081 & 155.12 & 0.6298 & 179.93 & 0.4285 & 36.42 & 0.2399 & 115.74 & 0.2975 \\
        {} & OmniGen \cite{xiao2025omnigen}  & 59.66 & 0.4596 & 46.18 & 0.4188 & 150.80 & 0.4663 & 119.87 & 0.3081 & 42.77  & 0.2169 & 134.43 & 0.3438\\
        {} & SuperEdit \cite{li2025superedit} & 89.07 & 0.5481 & 56.22 & 0.5307 & 172.50 & 0.5866 & 185.98 & 0.4489 & 49.27 & 0.2960 & 116.37 & 0.3860 \\
        {} & ICEdit \cite{zhang2025context} & 50.14 & 0.4922 & 45.54 & 0.4734 & 128.55 & 0.5385 & 149.44 & 0.3300 & 170.11 & 0.5961 & 104.72 & 0.2882 \\
        \midrule
        \multirow{3}{*}{\rotatebox{90}{Zero-IR}} & PSLD \cite{rout2023solving} & \textbf{31.90} & 0.2839 & 42.89 & 0.3683 & 115.17 & 0.3660 & - & - & - & - & 202.71 & 0.6242\\
        {} &TReg\cite{kim2023regularization}& 49.15 & 0.5161 & 52.07 & 0.4379 & 94.11 & 0.5392 & - & - & - & - & 183.27 & 0.7713\\
        {} &DAPS\cite{zhang2025improving} & 47.14 & 0.3290 & 59.85 & 0.3413 & 148.42 & 0.4137 & - & - & - & - & 213.36 & 0.6266 \\
        \midrule
        
        \multirow{3}{*}{\rotatebox{90}{IE-Edit}} & 
        VISII \cite{nguyen2023visual}& 110.39 & 0.4949 & 122.63 & 0.5465& 248.79 & 0.8341 & 203.83 & 0.5011 & 198.69 & 0.6756 & 298.10 & 0.5402 \\
        {} & Analogist \cite{gu2024analogist}& 83.88 & 0.4779 & 75.06 & 0.4692 & 143.62 & 0.5599 & 158.29 & 0.6006 & 68.02 & 0.3988 & 156.28 & 0.5779 \\
        {} & EditClip \cite{wang2025editclip}& 77.64 & 0.5558 & 78.75 & 0.5114 & 99.00 & 0.5470 & 174.93 &0.3809 & 44.69 & 0.2241 & 138.34 & 0.3008 \\
        \midrule
        \rowcolor{lightgray!20} \cellcolor{white} {}& One-Shot & 41.41 & 0.2666 & \textbf{35.51} & 0.2654 & 89.51 & 0.2801 & 87.12 & 0.2559 & 35.52 & 0.1633 & 107.70 & 0.2908\\
        \rowcolor{lightgray!20} \cellcolor{white} {}& Multi-Shot & 45.89 & 0.2654 & 42.62 & 0.2651 & 88.58 & 0.2846 & 69.52  & 0.2214 & 34.18 & 0.1584 & 107.80 &0.2744 \\
        \rowcolor{lightgray!20} \cellcolor{white!20} \multirow{-3}{*}{\rotatebox{90}{VDC}}& MS+Inverse-Correction & 45.00 & \textbf{0.2624} & 41.09 & \textbf{0.2593}& \textbf{82.57} & \textbf{0.2768} & \textbf{66.92} & \textbf{0.2155} & \textbf{33.23} & \textbf{0.1560} & \textbf{105.26} & \textbf{0.2729} \\
        \bottomrule
    \end{tabularx}

    \label{tab:bench}
    \vspace{-5mm}
\end{table*}

\subsection{Optimization and Inversion Refinement}

Previous condition optimization methods typically optimize the condition using the output of a single diffusion step~\cite{mokady2023null}.
However, this approach forces most edits to occur during the early diffusion steps, leaving the later stages primarily for refinement.
In contrast, our VDC optimizes all condition generators jointly based on the final output after the complete diffusion process.
This formulation allows the model to decide how edits are distributed across the diffusion trajectory, rather than concentrating them in the initial steps.
Accordingly, the optimization in \ref{eq:ours_single} becomes:
\vspace{-2mm}
\begin{flalign}
\nonumber &C^s_{p,...1} = \text{MLP}_{1,...p}(1,....N)\\
&\underset{\text{MLP}_{1,...p}}{\text{min}}|| Z_{0}^* - Z_{0}(Z_p,t=(p,...1),C_{p,...1}) ||_2^2
\end{flalign}
\vspace{-4mm}

Here, $N$ is the number of tokens in the condition, $p \sim [T,0)$ denotes the starting step of the partial diffusion process, and $t \sim [p,0)$ represents the current diffusion step.
$z_{0}^*$ is the ground-truth latent, while $z_{0}$ is the model output obtained using the optimized steering condition $C_{p,...1}$.
This formulation provides the model with greater flexibility to adapt the applied edits dynamically at each diffusion step.

\myparagraphnospace{Inversion Correction.}
DDIM inversion assumes that $Z_{t-1} \sim Z_t$, meaning that adjacent diffusion steps are nearly identical.
However, this assumption holds only for infinitesimally small step sizes, and in practice, it introduces accumulated inversion errors across the diffusion trajectory.
To improve inversion accuracy, we propose a refinement method for DDIM inversion.
We first perform DDIM inversion up to the desired diffusion step to obtain the initial noised latent $z_p$.
Next, we apply the forward diffusion process using the inverted latent to compute the reconstruction error.
Finally, we update the noised latent $z_p$ through gradient-based optimization to minimize this inversion error.
The full procedure is summarized in Algorithm~\ref{alg:ddim}.

\vspace{1mm}
\myparagraphnospace{Loss.}
Since our approach relies on visual examples, we convert the ground-truth image to the latent space and compute the loss directly in that domain.
This avoids the disparity between pixel and latent spaces~\cite{rout2023solving}, where multiple images may correspond to the same latent representation.
However, encoding an image into latent space can result in the loss of fine spatial details, producing inaccurate or overly smoothed edits.
To address this, we additionally compute a pixel-space loss by decoding the diffusion latent output back to the image domain.
Combining both latent and pixel losses helps preserve spatial fidelity while maintaining semantic consistency during editing:

%% file: sec/4_experiments.tex
\section{Experiments}

\begin{figure*}
    \centering
    \includegraphics[width=\textwidth]{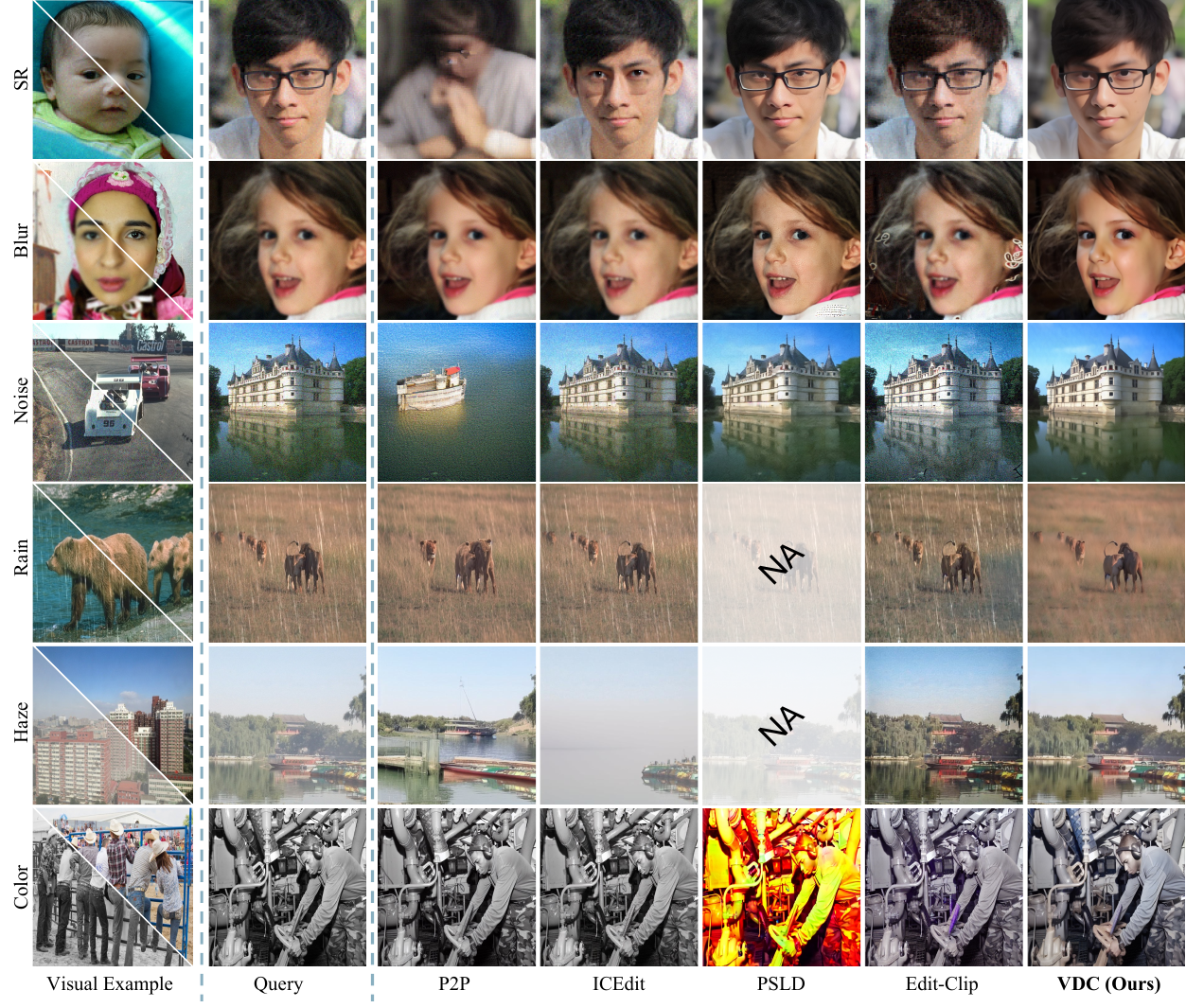}
    \vspace{-6mm}
    \caption{\textit{Visual comparison.} Text- and example-based methods struggle with complex edits due to misalignment or degradation priors. Our one-shot VDC (shown results) yields clean results, with multi-shot and correction modules improving generalization and fidelity.}
    \label{fig:exp:bench_visuals}
    \vspace{-4mm}
\end{figure*}
We conduct experiments across diverse editing and restoration tasks, comparing against works that adapt T2I diffusion models under different input modalities, training regimes, and optimization strategies.
For fairness, we use the same diffusion backbone and include instruction-based models explicitly trained for image editing.

\vspace{1mm}
\myparagraphnospace{Implementation Details.}
VDC builds on Stable Diffusion v1.4~\cite{rombach2022high} with DDIM sampling~\cite{song2020denoising} using 100 steps, operating only on the last 10 steps of the trajectory.
The condition generator (CG) is a three-layer MLP network with dimensionality $128$.
We optimize CG with Adam ($\beta_1{=}0.9$, $\beta_2{=}0.999$) for 200 iterations (batch size 4) using a cosine-annealed learning rate decaying from $5{\times}10^{-3}$ to $1{\times}10^{-3}$~\cite{loshchilov2016sgdr}.
The one-shot setup uses single visual example with flip, rotation, and color-jitter augmentations, while the multi-shot setup increases to eight examples.
Condition steering is set to a scale of 7.
All experiments run on a single RTX 4090 GPU. 
We use identical settings across architectures (e.g., SD~\cite{rombach2022high} vs. SANA~\cite{xie2024sana}) and tasks.

\vspace{1mm}
\myparagraphnospace{Datasets.}
For super-resolution and deblurring, we use 1K FFHQ~\cite{karras2019style} samples following DPS~\cite{chung2023diffusion} degradation.
We choose BSD400~\cite{arbelaez2010contour} testset $\sigma{=}25$ for denoising.
For deraining and dehazing, we evaluate on Rain100L~\cite{yang2020learning} and SOTS~\cite{li2018benchmarking}, respectively.
For colorization, we convert DIV2K~\cite{agustsson2017ntire} to grayscale.
We randomly pick one image per dataset as reference for works requiring visual examples.

\vspace{1mm}
\myparagraphnospace{Baselines.}
\textit{Text-edit} (T-Edit) methods manipulate the generation prompt without retraining. We use BLIP~\cite{li2023blip} to generate captions (\textit{e.g.}, “photo of 3 bears in rain” → “photo of 3 bears”) as editing prompts.
\textit{Instruction-edit} (I-Edit) methods are trained for text-instruction-based editing; we craft task-specific prompts (\textit{e.g.}, “Remove rain from the image”).
\textit{Zero-shot image restoration} (Zero-IR) methods address inverse problems using diffusion priors; we follow DPS~\cite{chung2023diffusion} for degradation settings.
\textit{Image-example} (IE-Edit) methods transfer edits from a reference image to a target; we use the same visual examples as our method for fair comparison.
Please refer to the supplementary for more details.

\vspace{-1mm}
\subsection{Comparison to State-of-the-Art Methods}

In~\cref{tab:bench}, VDC surpasses all prior approaches using only a \textit{single} visual example. Its language-free design provides stronger conditioning than text, overcoming the misalignment that limits text-based methods.
IE-Edit methods underperform due to their reliance on joint text–image embeddings, while Zero-IR methods perform better but require known degradation kernels, limiting generalization. Additionally, we compare to diffusion fine-tuning methods like ControlNet~\cite{zhang2023adding} and LoRA~\cite{hu2022lora} in \cref{tab:bench-sana}, showing their ineffectiveness under low-data regime.
Adding more visual examples further improves performance, especially for tasks with diverse degradations (see~\cref{sec:ablations}). VDC effectively captures multiple variations (e.g., rain patterns) within a single optimized condition, though too many examples may cause slight overfitting on less variable tasks.
Despite inverting only $10\%$ of the diffusion trajectory, the correction inversion module in the multi-shot setup improves detail preservation, particularly for deraining and denoising.
VDC is also efficient, requiring about 30 minutes of condition optimization for peak fidelity (200 optimization steps). However, Fig. \ref{fig:speeeed} shows that VDC outperforms OmniGen in just 10 steps ($\sim2$ mins). Additionally, Inference incurs zero overhead (just 10 timesteps), as VDC replaces CFG, leaving latency determined by the underlying diffusion model.
\textit{Please refer to supplementary for more results.}

\myparagraphnospace{Visual Results.}
Fig.~\ref{fig:exp:bench_visuals} illustrates that text-based methods (P2P, ICEdit) fail to perform complex edits due to text–visual misalignment, often producing corrupted results. Image-example approaches (Edit-CLIP) show similar issues, as they still depend on textual space.
Zero-IR methods generate cleaner outputs but introduce noise or color artifacts and rely on known degradation kernels, reducing their applicability to tasks such as deraining and dehazing.
In contrast, our one-shot VDC accurately captures task-specific visual features, achieving clean, artifact-free results.
As shown in Fig.~\ref{fig:exp:num_examples}, the multi-shot setup generalizes to more complex edits (e.g., colorization), while the correction inversion module further improves fidelity and consistency.
Together with the quantitative comparisons, these results showcase the effectiveness of our approach.

%% file: sec/4.2_Ablation.tex
\begin{table}[t]
    \centering
    \footnotesize
    \fboxsep0.75pt
    \caption{\textit{Contribution analysis.} The upper half evaluates the impact of each module, while the lower half compares different configurations of the condition generator (CG). \textbf{Best} results are bolded; the final setup is \colorbox{lightgray!20}{highlighted}.}
    \vspace{-3mm}
    \setlength\tabcolsep{2pt}
    \begin{tabularx}{\columnwidth}{lX*{5}{c}}
        \toprule
        {} & \multirow{2}{*}{Method} & \multicolumn{2}{c}{SR} & \multicolumn{2}{c}{DeRain} \\
         {} & {} & FID $\downarrow$ & LPIPs $\downarrow$ & FID $\downarrow$ & LPIPs $\downarrow$\\
        \midrule
        \multirow{4}{*}{\rotatebox{90}{Modules}} & w/o Data Augmentation & 48.53 & 0.2958 & 131.82 & 0.3352\\
        {} & w/o Pixel Loss & 46.93 & 0.2881 & 93.79 & 0.2723\\
        {} & w/o Condition Steering & 55.08 & 0.2726 & 122.20 & 0.26858\\ 
        {} & w/o Condition Generator & 44.31 & 0.2718 & 106.87 & 0.2568 \\
        \midrule
        {} & Single/Not-Conditioned & 41.74 & 0.3048 & 89.82 & 0.2567 & \\
        {} & Single/Step-Conditioned & 46.49 & 0.3135 & 93.67 & 0.2585 \\
        {} & Per-Step/Text-Conditioned & 56.88 & 0.3331 & 119.59 & 0.2828\\
        \rowcolor{lightgray!20} \cellcolor{white} \multirow{-4}{*}{\rotatebox{90}{CG-Setup}} & Per-Step/Not-Conditioned & \textbf{41.41} & \textbf{0.2666} & \textbf{87.12} & \textbf{0.2559} \\
        \bottomrule
    \end{tabularx}

    \label{tab:modules-ab}
    \vspace{-3mm}
\end{table}

\begin{figure}[t]
    \centering
    \includegraphics[width=\linewidth]{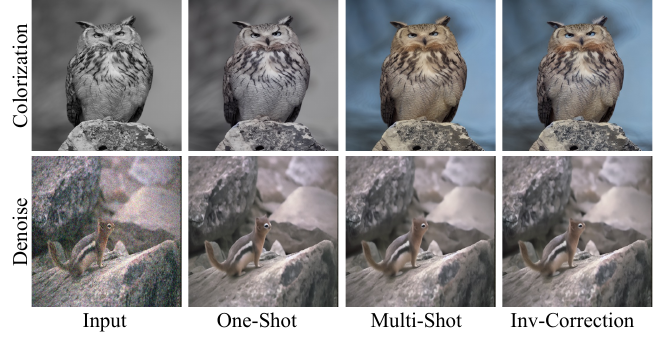}
    \vspace{-8mm}
    \caption{\textit{Number of visual examples.} Increasing the number of examples improves results, especially for tasks with high variability such as colorization. The inversion correction module further enhances detail preservation and overall output quality.}
    \label{fig:exp:num_examples}
    \vspace{-5mm}
\end{figure}

\vspace{-1mm}
\subsection{Ablations}
\label{sec:ablations}

We analyze the contribution of each component and design choice in our method, along with insights into diffusion behavior from a conditioning perspective. All experiments use the One-Shot setup on SR and Derain tasks; additional results are in the supplementary material.

\begin{figure}[t]
    \centering
    \includegraphics[width=\linewidth]{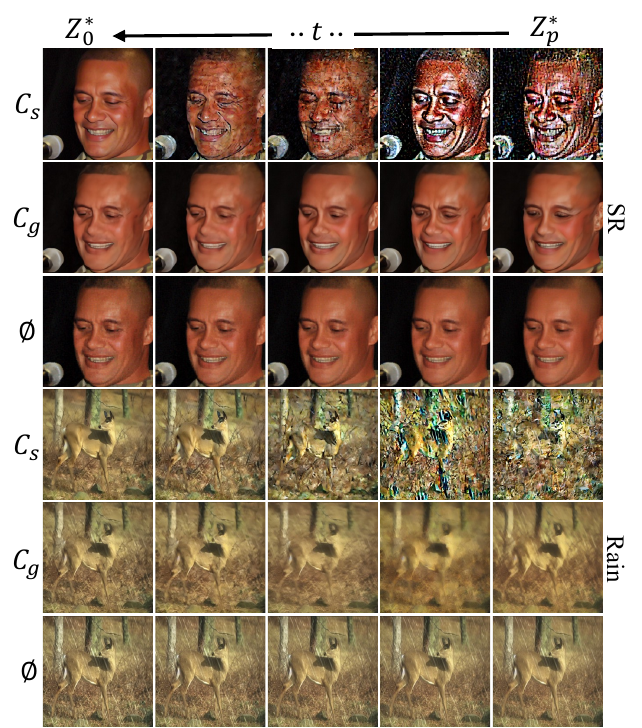}
    \vspace{-5mm}
    \caption{\textit{Condition Steering ($C_s$) vs. Condition Generation ($C_g$).} $C_s$ adapts the unconditional path $\phi$ for the target edit, whereas $C_g$ generates a new image from scratch.}
    \label{fig:abl:diff_path}
    \vspace{-5mm}
\end{figure}

\vspace{1mm}
\myparagraphnospace{Module Contributions.}
As shown in Tab.~\ref{tab:modules-ab}, we analyze the contribution of each proposed module. 
\textit{(I) Data augmentation} is crucial in the One-Shot setup, preventing overfitting to the single visual example and improving generalization across diverse patterns. 
\textit{(II) Pixel loss} substantially enhances quality, as relying solely on latent-space loss discards fine details that the model may misinterpret as edits. 
\textit{(III) Condition Generator (CG)} implemented as an MLP, improves stability and generalization by generating the full condition jointly rather than optimizing tokens independently. 
\textit{(IV) Condition Steering} provides the largest improvement by optimizing a steering condition that guides the unconditional diffusion trajectory toward the desired edit instead of generating a new image. This focuses the optimization on the edit itself, avoiding entanglement with example content and reducing artifacts. As shown in Fig.~\ref{fig:abl:diff_path}, our method effectively steers samples within the data distribution toward the target output, producing cleaner and more faithful edits.

\begin{figure}[t]
    \begin{subfigure}{0.45\columnwidth}
    \centering
        \includegraphics[width=\textwidth]{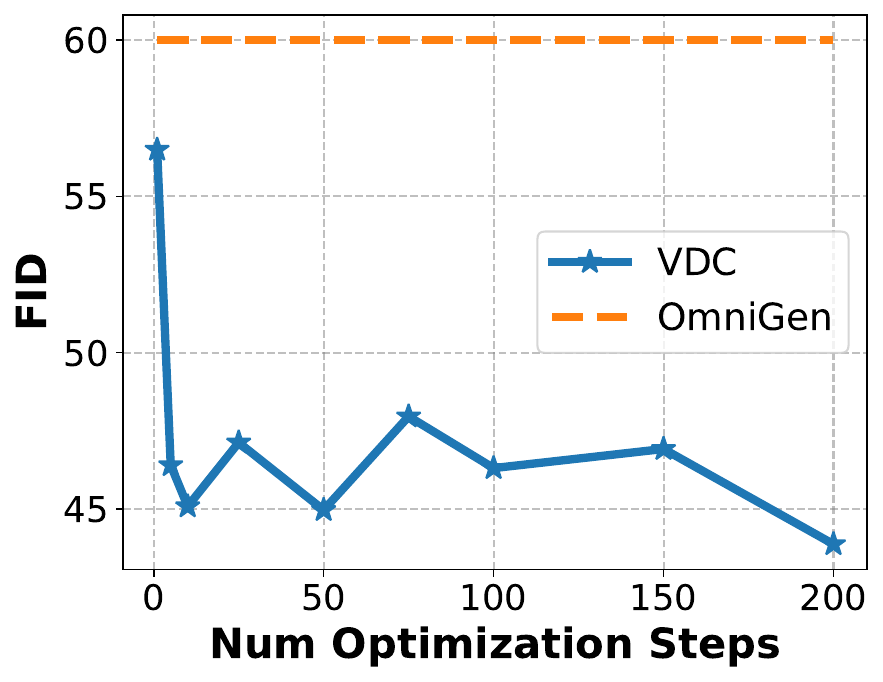}
        \subcaption{\textit{SR}}
    \end{subfigure}
    \hfill
    \begin{subfigure}{0.45\columnwidth}
    \centering
        \includegraphics[width=\textwidth]{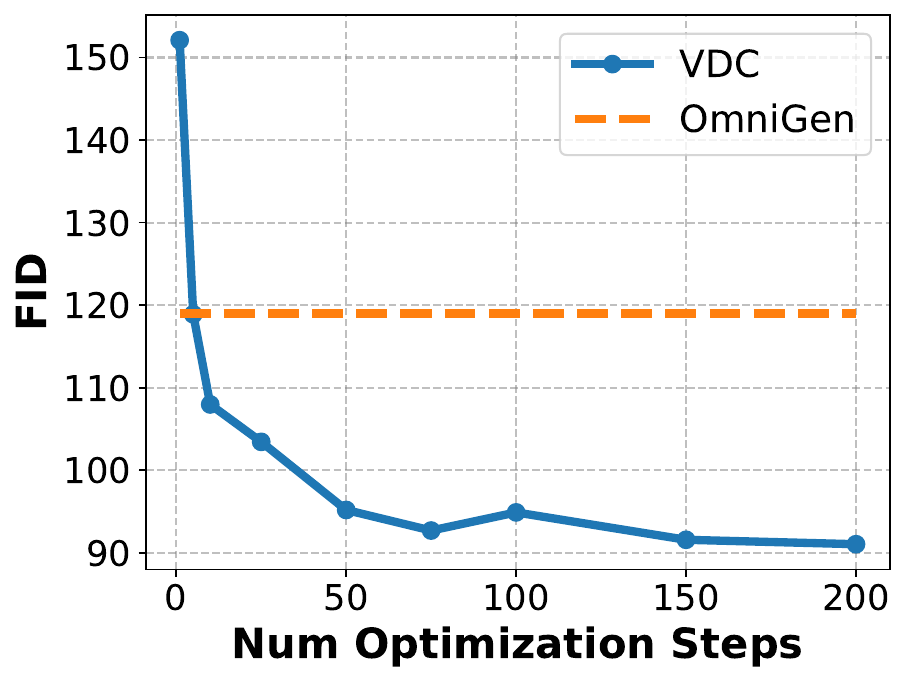}
        \subcaption{\textit{DeRain.}}
    \end{subfigure}
     \vspace{-3mm}
     \caption{\textit{Optimization trade-off}. VDC outperforms OmniGen in just 10 steps ($\sim2$m); extended optimization is optional.}
     \vspace{-5mm}
     \label{fig:speeeed}
\end{figure}

\vspace{1mm}
\myparagraphnospace{Condition Generator Setup.}
As shown in the lower half of Tab.~\ref{tab:modules-ab}, we evaluate different setups for condition generation. 
Using a separate condition for each sampling step increases the number of optimization parameters but grants the model greater flexibility, allowing step-specific updates that improve results. 
We implement this either by feeding the step index as input to a single generator or by assigning a dedicated generator to each step. The latter performs better, offering more expressive power and independence across steps. 
Initializing the generator with a text-based condition, however, reintroduces the text–visual misalignment problem, as the conditioning shifts back into textual space, leading to a notable performance drop. 
Our final setup employs independent generators for each diffusion step without any textual conditioning. 
Despite using multiple generators, the additional computational cost is negligible due to the small number of diffusion steps (10) and the compact size of each generator network (approx. 100K parameters).

\begin{figure}
    \centering
    \includegraphics[trim={3cm 1.5cm 3cm 2cm},clip, height=3cm, width=.8\linewidth]{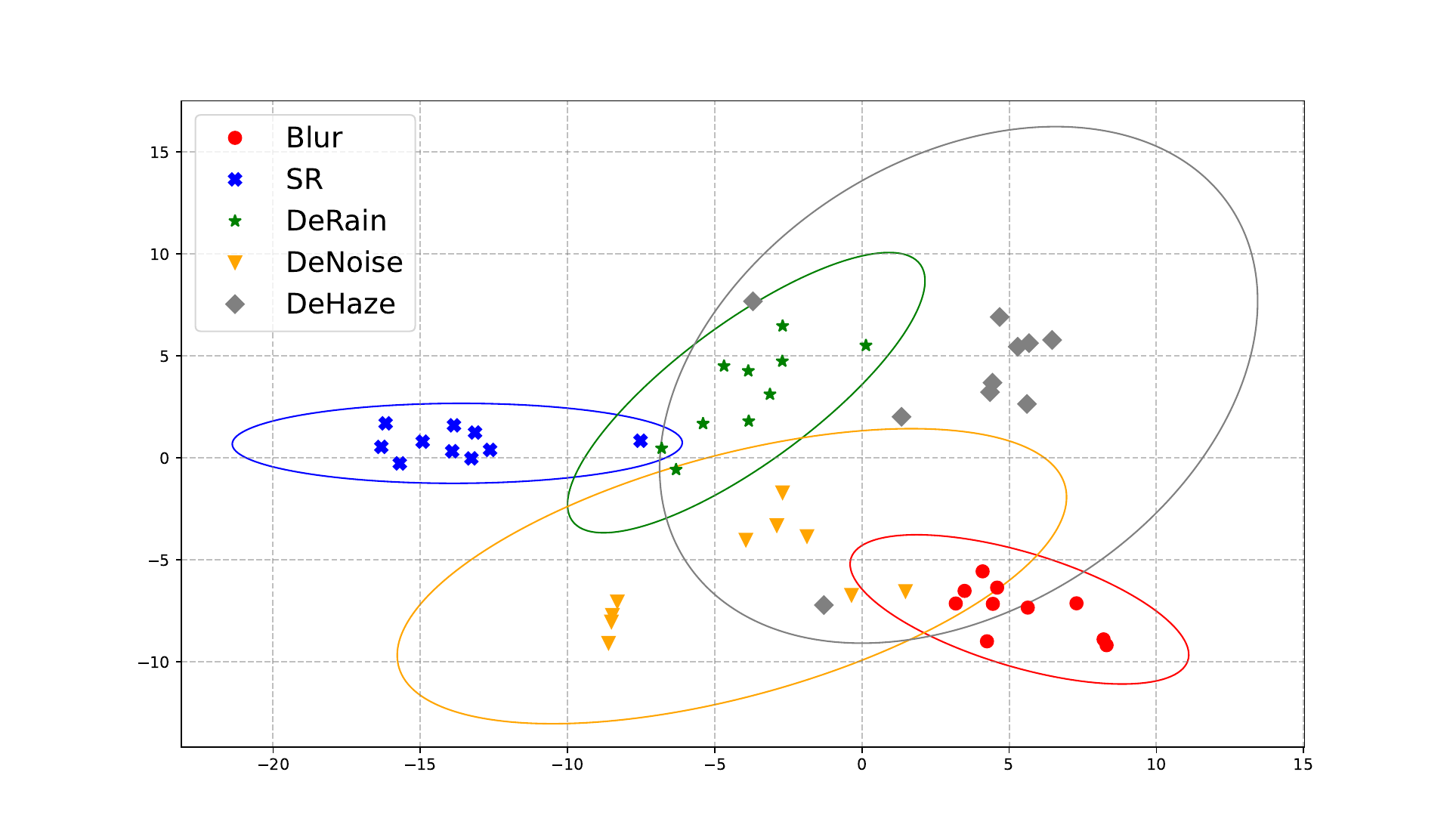}
    \vspace{-2mm}
    \caption{\textit{T-SNE visualization.} Conditions from the same task form clear clusters, showing that similar visual features (\textit{e.g.}, rain, blur) are recognized without textual dependency. This enables one-shot adaptation via condition optimization.}
    \label{fig:vis-conds}
    \vspace{-2mm}
\end{figure}

\vspace{1mm}
\myparagraphnospace{VDC captures Degradation Attributes.}
To analyze how visual conditions represent image degradations, we optimized a separate condition for 10 samples per task and visualized them in the condition space in Fig.~\ref{fig:vis-conds}. 
Conditions from the same task form compact clusters, indicating that similar degradations, such as rain or blur, are consistently encoded as related features, independent of textual representations. 
This confirms that our visually optimized conditions capture semantic similarities across varying appearances, enabling effective adaptation. Further, in Fig. \ref{fig:sensitivity}, we report the performance variance
across these models. Despite minor fluctuations in complex tasks (DeRain), performance remains robust regardless of the chosen example.

\begin{table}[t]
    \centering
    \footnotesize
    \fboxsep0.75pt
    \caption{\textit{Diffusion path length.} Extending the diffusion path increases variation at the cost of fidelity. \textbf{Best} results are bolded; the final setup is \colorbox{lightgray!20}{highlighted}.}
    \setlength\tabcolsep{8pt}
    \vspace{-3mm}
    \begin{tabularx}{\columnwidth}{X*{4}{c}}
        \toprule
         \multirow{2}{*}{Path Length} & \multicolumn{2}{c}{SR} & \multicolumn{2}{c}{DeRain} \\
          {} & FID $\downarrow$ & LPIPS $\downarrow$ & FID $\downarrow$ & LPIPS $\downarrow$\\
        \midrule
        5\% & 56.86 & 0.3037 & 89.89 & \textbf{0.2470}\\
        \rowcolor{lightgray!30} 10\% & \textbf{41.41} & \textbf{0.2666} & \textbf{87.12} & 0.2559 \\
        20\% & 56.26 & 0.3134 & 104.33 & 0.2708 \\
        30\% & 58.12 & 0.3193 & 107.64 & 0.2856 \\
        
        \bottomrule
    \end{tabularx}
    \label{tab:stength-ab}
    \vspace{-6mm}
\end{table}

\begin{figure}
    \centering
    \includegraphics[width=\linewidth]{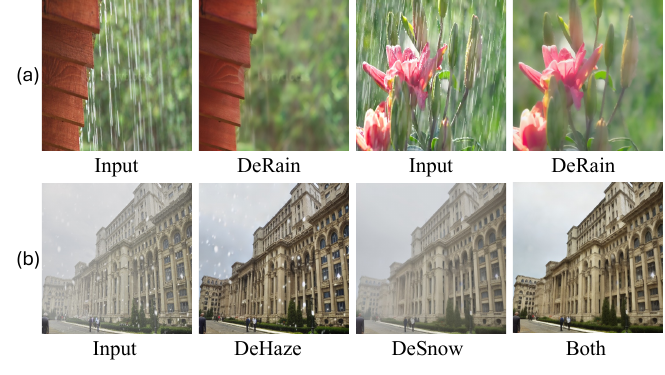}
    \vspace{-7mm}
    \caption{\textit{Generalization $\&$ expressiveness.} (a) VDC generalizes from synthetic to real data (RealRain-1K~\cite{li2022toward}). (b) Visual examples enable fine-grained edits (CDD-11~\cite{guo2024onerestore}).}
    \label{fig:generalization}
    \vspace{-2mm}
\end{figure}

\vspace{1mm}
\myparagraphnospace{Generalization and Expressiveness.}
Our method not only learns from synthetic examples but also generalizes to real data and diverse editing scenarios, highlighting the flexibility and scalability of visual conditioning.
\textit{(I) Generalization to real data.} Leveraging diffusion priors, VDC generalizes beyond synthetic data, performing well on real images—only eight synthetic samples enable effective deraining on unseen rain patterns (\cref{fig:generalization}\textcolor{cvprblue}{a}).
\textit{(II) Expressiveness}. Visual examples offer more precise and controllable conditioning. As shown in~\cref{fig:generalization}\textcolor{cvprblue}{b}, they clearly separate degradations (\textit{e.g.}, haze vs.\ snow), whereas text prompts often blur this distinction.
\textit{(III) Generality.} VDC is model-agnostic and applicable to any conditional diffusion framework, including flow-matching models~\cite{lipman2022flow}. Since it learns edits directly from visual examples, it naturally extends beyond pixel-level tasks to broader semantic and object-level modifications.
\textit{(IV) Multi-tasking.}~\cref{fig:generalization}\textcolor{cvprblue}{b} proves a single embedding can learn concurrent tasks (e.g., DeSnow+DeHaze), consolidating multiple needs into one ”generalist” solution.

\vspace{1mm}
\myparagraphnospace{Diffusion Path Length.}
Following deeper into the diffusion trajectory introduces greater noise to the latent, expanding the output space but increasing deviations from the input image.
As visualized in Tab.~\ref{tab:stength-ab}, using longer diffusion paths degrades performance, while overly short paths limit the reachable output space and prevent the desired edit.

%% file: sec/5_conclusion.tex
\section{Conclusion}

Text-guided diffusion models remain limited by weak language–vision alignment, hiding much of their visual editing potential behind text conditioning.
\textit{Visual Diffusion Conditioning} (VDC) unlocks this potential by replacing text with visual examples as the source of guidance.
VDC learns visual conditions directly from paired examples and steers the diffusion process toward precise, language-free edits through a lightweight condition generator and a condition-steering mechanism.
An inversion correction step further preserves fine details and realism.

With as little as one example, VDC adapts text-to-image diffusion models for complex edits such as deraining, deblurring, and dehazing—without retraining or fine-tuning.
It achieves accurate, artifact-free results while remaining efficient, training-free, and generalizable to real-world data.
Future work could explore extending VDC to unposed or in-the-wild images and studying its behavior on more diverse real-world conditions.

\noindent \textbf{Acknowledgments:}  This work was supported by the Alexander von Humboldt Foundation.

%% file: sec/supp.tex
\clearpage

\setcounter{section}{0}
\setcounter{figure}{8}
\setcounter{table}{3}

\renewcommand{\thesection}{\Alph{section}}

\maketitlesupplementary
In the supplementary material, we first provide further details of the compared prior works in \cref{sec:supp:priorworks}. 
Additional ablations and analyses on out-of-distribution tasks are presented in \cref{sec:supp:ood_ablations}, and \cref{sec:supp:hyper} examines the sensitivity of VDC to its hyperparameters. 
We then analyze the computational complexity of VDC in \cref{sec:supp:complexity}. 
Finally, \cref{sec:supp:limitations} discusses the limitations of our approach, and \cref{sec:supp:visuals} includes extended visual comparisons for all compared methods.

\section{Further Details on Prior Works}
\label{sec:supp:priorworks}
We provide additional details on prior works used for comparison, highlighting their reliance on language.

\vspace{1mm}
\myparagraphnospace{Text-Prompt Editing Methods (T-Edit).}
These methods require a complete text description of the input image. We use this description as the text prompt to condition both the inversion and generation processes. To enable each edit, we manually append the visual attribute corresponding to the target task. Captions are generated using BLIP~\cite{li2023blip}.

The text prompt for each degradation type is constructed as follows: \textit{(i)} Rain: ``[Text Description] + in the rain'', \textit{(ii)} Fog: ``[Text Description] + in the fog'', \textit{(iii)} SR: ``Low-resolution image of [Text Description]'', \textit{(iv)} Blur: ``Blurry image of [Text Description]'', \textit{(v)} Noise: ``Noisy image of [Text Description]'', \textit{(vi)} Colorization: ``Grayscale image of [Text Description]''

\begin{itemize}
\item \textbf{Prompt-to-Prompt (P2P)} \cite{hertz2022prompt}: Manipulates cross-attention during generation to adjust visual features associated with specific prompt words. For our tests, we mask cross-attention features tied to the degradation being removed (e.g., rain, fog, noise).
\item \textbf{Null-Text Optimization (Null-Opt)} \cite{mokady2023null}: Improves DDIM inversion for image editing. We apply this optimization jointly with P2P for all edits.
\item \textbf{Negative Condition} \cite{miyake2025negative}: Replaces standard null-text conditioning in classifier-free guidance with negative prompts describing the unwanted degradation (e.g., "fog, foggy, haze, hazy, blurry, blur" for dehazing; "noise, noisy, low quality'' for denoising).
\end{itemize}

\vspace{1mm}
\myparagraphnospace{Text-Instruction Editing (I-Edit.)}
These methods take an input image and a natural-language instruction describing the desired edit. Each model is trained or fine-tuned to apply edits according to the instruction. We use the default configurations provided in the authors’ open-source implementations.

For text instruction, we used: for DeRain "Remove rain and water drops from the image", for DeHaze "Remove fog and haze from the image", for SR "Increase image resolution, improve quality and remove noise", for DeBlur "Increase image sharpness, improve quality and remove noise" for DeNoise "Remove noise from the image", for Colorization "Color this grayscale image".

\vspace{1mm}
\myparagraphnospace{Zero-Shot Image Restoration (Zero-IR).}
Zero-IR methods solve inverse problems using diffusion models as strong generative priors. They require a degradation kernel that models the corruption in the input image. These methods search the diffusion latent space for an image that degrades to an image that matches the input. We use the released code and default task-specific settings for each method. For colorization, we adopt the kernel settings from Zero-Null~\cite{wang2022zero}. For the remaining tasks, we use the kernels defined in DPS~\cite{chung2023diffusion}.

\vspace{1mm}
\myparagraphnospace{Image Exemplar-based Editing (IE-Edit).}
These methods infer an edit from a before/after image pair and apply that edit to a new image. For fair comparison, we use the same reference example images employed to optimize our method.

\begin{itemize}
\item \textbf{VISII} \cite{nguyen2023visual}: Builds on the text-instruction editing model Instruct-Pix2Pix~\cite{brooks2023instructpix2pix}. It optimizes a text instruction that reproduces the edit shown in the example pair, then applies the resulting instruction to new inputs.
\item \textbf{Analogist} \cite{gu2024analogist}: Uses Stable Diffusion Inpainting~\cite{rombach2022high} together with a large language model that extracts the transformation between example images. It then applies this transformation to new inputs via inpainting.
\item \textbf{EditClip} \cite{wang2025editclip}: Fine-tunes the CLIP image encoder~\cite{radford2021learning} to capture relationships between the example images. It further fine-tunes Stable Diffusion~\cite{rombach2022high} to condition on these relationships. The model is trained on hundreds of thousands of edited images paired with text instructions.
\end{itemize}

\begin{table*}[t]
    \centering
    \footnotesize
    \caption{\textit{VDC compared to fine-tuning and with different generative models.} FID ($\downarrow$) and LPIPS($\downarrow$) are reported on the full RGB images. Our method highly surpasses diffusion fine-tuning methods in low data regime. Additionally, VDC can be utilized with different conditional generative models. The \textbf{best} performances are highlighted.}
    \vspace{-3mm}
    \setlength\tabcolsep{3.4pt}
    \begin{tabularx}{\textwidth}{cX*{13}{c}}
        \toprule
        \multirow{2}{*}{Type} & \multirow{2}{*}{Method} & Num. & \multicolumn{2}{c}{SR} & \multicolumn{2}{c}{DeBlur} & \multicolumn{2}{c}{DeNoise} & \multicolumn{2}{c}{DeRain} & \multicolumn{2}{c}{DeHaze} & \multicolumn{2}{c}{Colorization} \\
        {} & {} & Samples& FID $\downarrow$ & LPIPS $\downarrow$ & FID $\downarrow$ & LPIPS $\downarrow$ & FID $\downarrow$ & LPIPS $\downarrow$ & FID $\downarrow$ & LPIPS $\downarrow$ & FID $\downarrow$ & LPIPS $\downarrow$ & FID $\downarrow$ & LPIPS $\downarrow$ \\
        \midrule
        \multirow{2}{*}{\rotatebox{90}{F-T}}& ControlNet~\cite{zhang2023adding} & 200& 110.62 & 0.4291 & 80.03 & 0.3225 & 145.01 & 0.4777 & 153.29 & 0.4803 & 50.00 & 0.2750 & 143.50 & 0.4093\\
        {}& PairEdit~\cite{lu2025pairedit} & 8 & 98.68 & 0.3134 & 67.08 & 0.3959 & 168.68 & 0.4561 & 148.59  & 0.3924 & 85.69 & 0.3924 & 150.32 & 0.5089 \\
        \midrule
        \multirow{3}{*}{\rotatebox{90}{SD}}& One-Shot & 1 & \textbf{41.41} & 0.2666 & 35.51 & 0.2654 & 89.51 & 0.2801 & 87.12 & 0.2559 & 35.52 & 0.1633 & 107.70 & 0.2908\\
        {}& Multi-Shot & 8 & 45.89 & 0.2654 & 42.62 & 0.2651 & 88.58 & 0.2846 & 69.52  & 0.2214 & 34.18 & 0.1584 & 107.80 &0.2744 \\
        {}& MS+Inv-Correc & 8 & 45.00 & \textbf{0.2624} & 41.09 & \textbf{0.2593}& \textbf{82.57} & \textbf{0.2768} & \textbf{66.92} & \textbf{0.2155} & \textbf{33.23} & \textbf{0.1560} & \textbf{105.26} & \textbf{0.2729} \\
        \midrule

        \multirow{3}{*}{\rotatebox{90}{SANA}} & 
        One-Shot & 1 & 50.20 & 0.2900 & 40.33 & 0.2587 & 82.72 & 0.2510 & \textbf{93.61}& 0.24807 & 29.20 & 0.1414 &107.74 & 0.26254  \\
        {} & Multi-Shot& 8 & 48.25 & \textbf{0.2478} & 33.99&0.2140 & 73.57&0.2485  & 98.80 & \textbf{0.2432} & 29.46 & 0.1403 & \textbf{104.97}& \textbf{0.2596} \\
        {} &  MS+Inv-Correc & 8 & \textbf{45.81} &  0.24834 & \textbf{32.38}& \textbf{0.2134} &	\textbf{69.65}  & \textbf{0.24816} & 97.54 & 0.2446 & \textbf{28.70} & \textbf{0.1398}	& 105.13 & 0.2603	\\
        \bottomrule
        
    \end{tabularx}
    \label{tab:bench-sana}
\end{table*}

\begin{table}[t]
    \centering
    \footnotesize
    \fboxsep0.75pt
    \caption{\textit{OOD Generalization.} We compare our method to state-of-the-art All-in-One Image Restoration (IR) on real image DeRain. We utilize RealRain-1k-L \cite{li2022toward} dataset for testing. Our method is able to generalize to real data while prior works fail. \textbf{Best} results are highlighted.}
    \setlength\tabcolsep{6pt}
    \vspace{-3mm}
    \begin{tabularx}{\columnwidth}{X*{4}{c}}
        \toprule
          Methods & Instruct-IR \cite{conde2024instructir} & MoCE-IR \cite{zamfir2025complexity} & VDC (\textit{ours})  &  \\
        \midrule
         FID $\downarrow$ & 124.82 &  154.94 & \textbf{106.89}\\
         LPIPS $\downarrow$ & 0.2553  & 0.3646 & \textbf{0.2154}\\
        
        \bottomrule
    \end{tabularx}
    \label{tab:OOD-ab}

\end{table}

\begin{figure}[t]
    \centering
    \includegraphics[width=\linewidth]{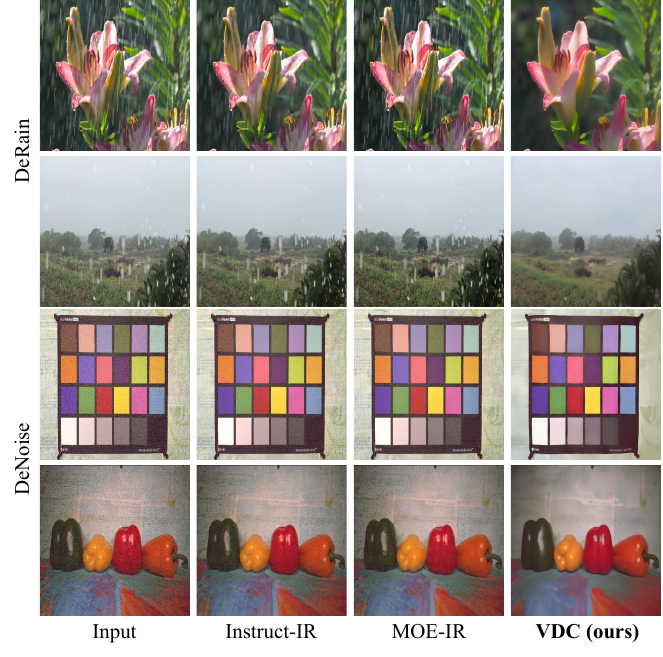}
    \caption{\textit{Visual comparison for OOD samples.} We compare our method to SOTA restoration models on real data. Our method is able to work on real data, whereas IR methods trained on syntactic data fail to generalize. We utilize RealRain-1k-L \cite{li2022toward} for Derain, and SIDD \cite{abdelhamed2018high} for denoising.}
    \label{fig:real-comp}
    \vspace{-3mm}
\end{figure}

\section{Ablations on Generalization}
\label{sec:supp:ood_ablations}
\subsection{VDC is model-agnostic}
To demonstrate that our method is not tied to a specific generative model and can generalize to any conditional generative framework, we adapt the SANA model~\cite{xie2024sana}—a conditional generative model based on Flow Matching~\cite{lipman2022flow}—for image editing using VDC. As shown in Tab.~\ref{tab:bench-sana}, VDC successfully enables SANA to perform image editing and restoration, and even surpasses the Stable Diffusion (SD)–based version on several tasks (DeNoise and DeHaze), benefiting from SANA’s more advanced generative prior. These improvements stem from SANA’s stronger prior, which yields higher-quality reconstructions of the inverted image.
However, SANA’s latent encoder applies a significantly higher compression rate (8× for SD versus 32× for SANA), which can limit the preservation of fine details in the latent space—an effect visible in the DeRain results. Our inversion correction module is likewise model-agnostic and can be used with Flow Matching models. As shown, it consistently improves performance, particularly on tasks that rely heavily on detail preservation.
Finally, the visual comparisons in Figs.~\ref{fig:exp:sr}–\ref{fig:exp:color} further illustrate that VDC successfully adapts SANA for high-quality image editing and restoration.

\subsection{VDC improves over Fine-Tuning}
Fine-tuning (F-T) and diffusion adaptation methods like ControlNet \cite{zhang2023adding} rely on massive supervision. Tab. \ref{tab:bench-sana} shows that fine-tuning fails in the low-data regime: ControlNet \cite{zhang2023adding} trained on 200 samples suffers from severe domain shift. Even a few-shot fine-tuning method like PairEdit \cite{lu2025pairedit}, based on LoRA and fine-tuned on 8 samples, yields poor fidelity. Additionally, it requires optimizing a new content LoRA for each inference image, requiring around 20-30 minutes of inference time. In contrast, VDC achieves strong results using only a single example and with zero inference overhead.

\subsection{Out-of-distribution performance}

A key advantage of adapting generative models for editing is the ability to leverage their real-data generative priors, enabling strong generalization to out-of-distribution inputs. In our framework, the generative model itself performs the edit—VDC simply provides a mechanism to communicate the desired transformation. In contrast, task-specific restoration or editing models learn the edit directly from training data, making their performance heavily dependent on the distribution and realism of that data. As a result, models trained on synthetic degradations often struggle to generalize to real-world scenarios.
Despite using only synthetic examples to optimize the steering condition, our method generalizes effectively to real data. As shown in Tab.~\ref{tab:OOD-ab}, VDC achieves strong real-world DeRain performance using just eight synthetic examples, successfully handling rain patterns that differ substantially from those in the examples. Meanwhile, specialized restoration models fail to generalize even when trained on large-scale synthetic datasets.

As shown in Fig.~\ref{fig:real-comp}, the gap between synthetic and real rain patterns causes traditional image restoration methods to fail at detecting and removing real rain streaks. In contrast, our approach leverages the generative model’s priors to correctly identify and remove these streaks, resulting in accurate edits. A similar trend is observed on real-world denoising data, where our method continues to generalize effectively while baseline restoration methods struggle.

\subsection{Performance on general editing tasks}
We center our benchmark on fine-detail edits, global adjustments, and image restoration tasks—categories where existing methods often struggle due to visual–text misalignment. Nonetheless, our approach is a general editing framework: it extracts the transformation from a given example and applies it to a new input. 

\begin{figure}[t]
    \centering
    \includegraphics[width=\linewidth]{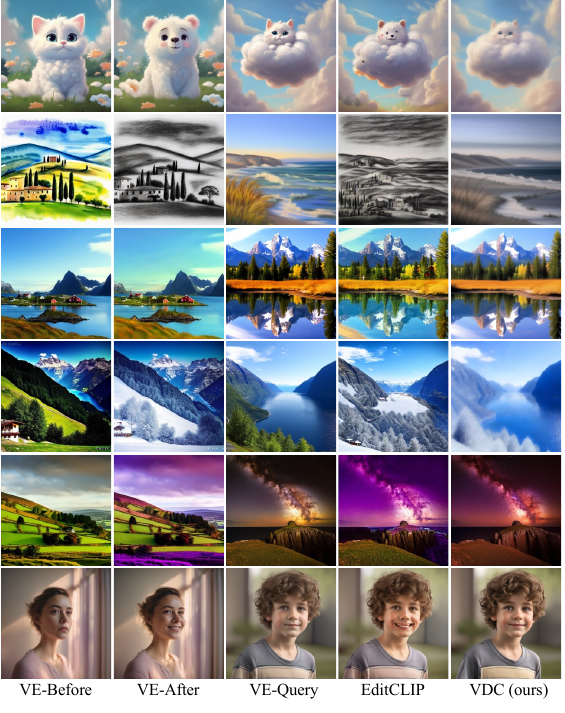}
    \vspace{-8mm}
    \caption{General Image Editing. We show the output of our method on general edits. Our method is not just limited to fine details or global edits but can also extend to semantic and object-specific edits. Images from  TOP-Bench \cite{zhao2024instructbrush} Dataset. }
    \label{fig:general-edits}
    \vspace{-3mm}
\end{figure}

As illustrated in Fig.~\ref{fig:general-edits}, by simply increasing diffusion path length (60\%), our method supports a wide range of edits, including semantic and object-specific modifications, compared to EditCLIP~\cite{wang2025editclip}, which is trained for visual-instruction–guided editing. Our method more reliably interprets the edits present in the example pair, particularly for global adjustments. EditCLIP may introduce unintended artifacts because its behavior is influenced by CLIP representation abilities and the common patterns in its large training corpus.

\begin{figure}[t]
     \centering

     \setlength{\tabcolsep}{1pt}
     \begin{tabular}{c c c c c c}
          \includegraphics[width=0.24\linewidth]{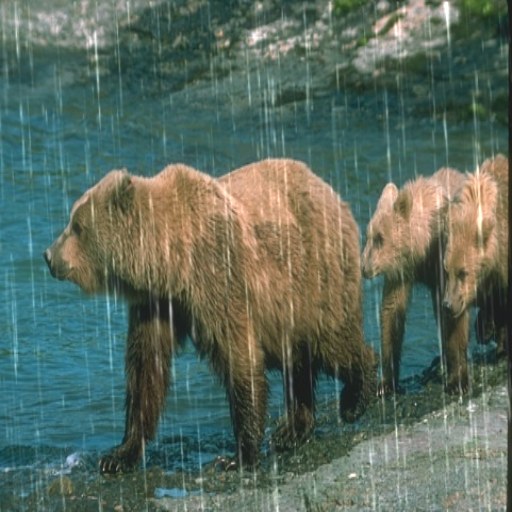} &
          \includegraphics[width=0.24\linewidth]{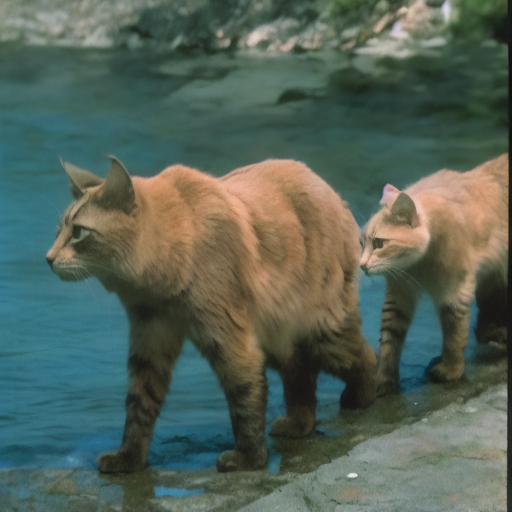} &
          \includegraphics[width=0.24\linewidth]{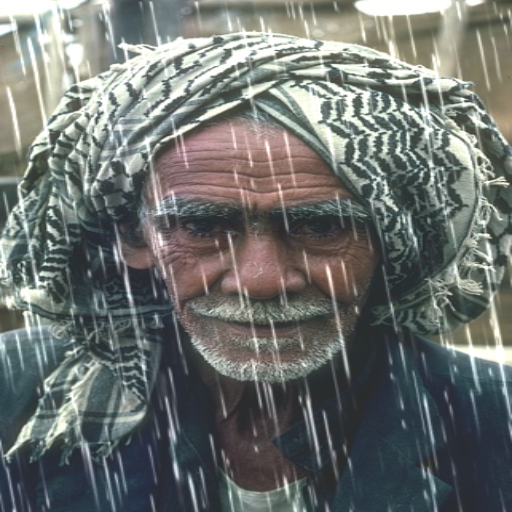} &
          \includegraphics[width=0.24\linewidth]{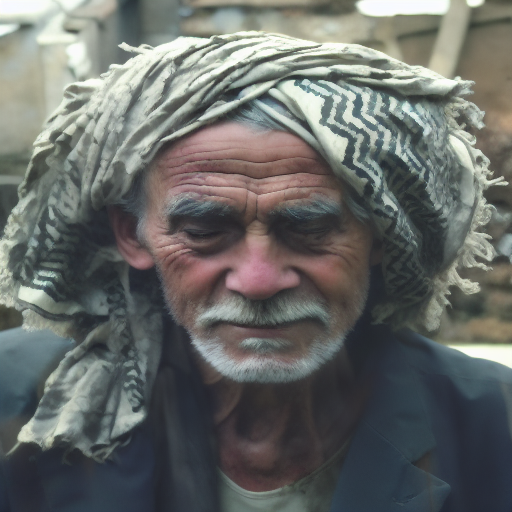} \\
          \footnotesize Input & \footnotesize + "Cats" & \footnotesize Input & \footnotesize + "Eyes Closed" \\
          
     \end{tabular}
     \vspace{-4mm}
    \caption{
    \textit{Composability of VDC.} VDC is used to steer the visual style while text independently controls the semantic content. 
}
     \vspace{-3mm}
     \label{fig:presets-img-samples}
\end{figure}

\vspace{1mm}
\myparagraphnospace{Semantic $\&$ Non-Rigid Edits.}
We clarify that VDC targets visual attribute steering (e.g., restoration, stylization) where text is ambiguous. 
To ensure high fidelity, we rely on pixel-space losses, which inherently prioritize structural preservation over non-rigid flexibility (e.g., pose changes).
However, VDC resolves this by supporting textual control: as shown in \cref{fig:presets-img-samples}, VDC handles visual patterns (DeRain) while text drives semantic shifts (e.g., bears$\to$cats) and non-rigid edits (e.g., closing eyes).

\section{Complexity Analysis}
\label{sec:supp:complexity}
As shown in Tab.~\ref{tab:complex}, the complexity of other methods is largely determined by the inference requirements of their underlying generative models. Zero-IR methods, in particular, incur significantly higher cost due to their sampling-based search procedures. In contrast, by directly optimizing the steering condition for the chosen sampling path, our approach minimizes the number of required inference steps. This yields the highest efficiency among the compared methods, requiring only 20 total steps for editing (10 DDIM inversion steps and 10 sampling steps) while still achieving the best performance. When the inversion correction module is used, the total number of steps increases, but this module is optional and can be enabled based on the task or available computational resources.

Although our method is entirely train-free, it still requires optimizing the steering condition for each adapted task. This optimization consists of 200 full-path iterations (2000 diffusion steps) and needs to be performed only once per task. On an RTX 4090 GPU, this process takes roughly 30 minutes. This is comparable to other train-free methods that rely on test-time optimization—such as Null-Opt~\cite{mokady2023null} (500 diffusion steps) and VISII~\cite{nguyen2023visual} (5000 diffusion steps)—but with the advantage that our optimization is performed per task rather than per inference. Overall, train-free approaches remain substantially more efficient than methods that require training or fine-tuning on hundreds of thousands of images across multiple GPUs for several days.

\begin{table*}[t]
    \centering
    \footnotesize
    \caption{\textit{User Study.} This table represents the preferences of the participants in the user study from the compared methods' outputs across different aspects. We report the choice percentage averaged across participants. ‘-’ represents unreported results. \textbf{Best} results are bolded.}
    \setlength\tabcolsep{3pt}
    \vspace{-3mm}
    \begin{tabularx}{\textwidth}{X*{8}{c}}
        \toprule
         \multirow{2}{*}{Method} & \multicolumn{4}{c}{SR} & \multicolumn{4}{c}{DeRain} \\
          {} &  Perceptual $\%$ & Artifact-Free $\%$ & Preservation $\%$ & Overall $\%$ & Perceptual $\%$ & Artifact-Free $\%$ & Preservation $\%$ & Overall $\%$\\
        \midrule
        Negative-Cond \cite{miyake2025negative} & 1 & 3.5 & 6 & 1 & \textbf{35} & 23.5 & 20.5 & 18.5\\
        OmniGen \cite{xiao2025omnigen} & 0.5 & 4.5 & 22 & 0.5 & 26.5 & 26.5 & \textbf{41.5} & 20\\
        PSLD \cite{rout2023solving} & \textbf{67}& 23.5 & 33.5 & 47.5 &-&-&-&-\\
        EditClip \cite{wang2025editclip} & 0 &	0 & 3.5 & 0 & 4.5 & 1 & 10.5 & 0\\
        VDC & 31.5 & \textbf{68.5}& \textbf{35}& \textbf{51} & 34 & \textbf{47.5} & 27.5 & \textbf{61.5}\\
        
        \bottomrule
    \end{tabularx}
    \label{tab:subj}
    \vspace{-3mm}
\end{table*}

\section{User Study}

To better evaluate the tested methods, the mean opinion score (MOS) was calculated through a user study by asking the participants to choose their favorite output according to different criteria. For a better understanding of the underlying task and accurate evaluation, the chosen user study participants are 10 imaging experts, including professional photographers. We conducted the user study on 2 different tasks (SR and DeRain) utilizing 20 samples from each task, selected randomly, and were fixed for all participants. We hide the names of the methods and randomly shuffle their position in the comparison grid to eliminate method bias. The comparison includes 5 different methods chosen by selecting the best overall performing method for its own type (Text Edit, Instruction Edit, etc). Our VDC method compared is the one-shot method optimized only on one visual example. We asked participants to choose their favorite output for four different categories: Best Perceptual Quality, Least Artifacts, Best Content Preservation, and lastly Best Overall for the task. MOS for all categories is represented in~\cref{tab:subj}. As we see from the results, our method is the most preferred by the users, with 51\% and 61\% choice as the preferred method in SR and DeRain, respectively. In SR PSLD \cite{rout2023solving} produces sharper images, which results in a higher perceptual quality; however, this method produces noticeable artifacts and noise (Fig.~\ref{fig:exp:sr},~\ref{fig:exp:blur}), resulting in our method being chosen as the best for the task. For DeRain, we notice a similar trend; however, because the other methods tend to create artifacts and content changes to the input, our method is still preferred by a big margin. We can appreciate our method's consistency across different tasks and compared aspects.

\begin{table}[t]
    \centering
    \footnotesize
    \caption{\textit{Complexity Analysis.} NFEs $\downarrow$ are Neural Function Evaluations. Our method sets a new state-of-the-art while being the most inference-efficient.  ‘-’ represents unreported results. The \textbf{best} performances are highlighted. }
    \vspace{-3mm}
    \setlength\tabcolsep{3.5pt}
    \begin{tabularx}{\columnwidth}{cX*{6}{c}}
        \toprule
        Type & Method & Train-Free & NFEs & Deblur & DeRain \\
        
        \midrule
        \multirow{3}{*}{\rotatebox{90}{T-Edit}} & P2P \cite{hertz2022prompt} & $\checkmark$ & 100 & 45.62 & 139.19\\
        {} & Null-Opt \cite{mokady2023null} & $\checkmark$ & 600 & 51.89 & 167.61\\
        {} & Negative-Cond \cite{miyake2025negative} & $\checkmark$ & 100 & 43.61 & 96.19\\
        \midrule
        \multirow{4}{*}{\rotatebox{90}{I-Edit}} & Instruct-Pix2Pix \cite{brooks2023instructpix2pix}& $\times$ & 100 & 142.91 & 179.93  \\
        {} & OmniGen \cite{xiao2025omnigen}& $\times$  & 50 & 46.18 & 119.87 \\
        {} & SuperEdit \cite{li2025superedit}& $\times$ & 100 & 56.22 & 185.98 \\
        {} & ICEdit \cite{zhang2025context}& $\times$ & 28 & 45.54 & 149.44 \\
        \midrule
        \multirow{3}{*}{\rotatebox{90}{Zero-IR}} & PSLD \cite{rout2023solving} & $\checkmark$ & 1000 & 42.89 & -\\
        {} &TReg\cite{kim2023regularization}& $\checkmark$ & 200 & 52.07 & -\\
        {} &DAPS\cite{zhang2025improving} & $\checkmark$ & 150 & 59.85 & - \\
        \midrule
        
        \multirow{3}{*}{\rotatebox{90}{IE-Edit}} & 
        VISII \cite{nguyen2023visual} & $\checkmark$ & 40 & 122.63 & 203.83 \\
        {} & Analogist \cite{gu2024analogist} & $\checkmark$ & 50 & 75.06 & 158.29\\
        {} & EditClip \cite{wang2025editclip} & $\times$ & 50 & 78.75& 174.93 \\
        \midrule
        \rowcolor{lightgray!20} \cellcolor{white} {}& One-Shot & $\checkmark$ & \textbf{20} & \textbf{35.51} & 87.12\\
        \rowcolor{lightgray!20} \cellcolor{white} {}& Multi-Shot & $\checkmark$ & \textbf{20} & 42.62 & 69.52\\
        \rowcolor{lightgray!20} \cellcolor{white} \multirow{-3}{*}{\rotatebox{90}{VDC}}& MS+Inverse-Correction & $\checkmark$ & 220 & 41.09 & \textbf{66.92} \\
        \bottomrule
    \end{tabularx}
    \label{tab:complex}
\end{table}

\section{Ablations on Hyperparameters}
\label{sec:supp:hyper}

\begin{table}[t]
    \centering
    \footnotesize
    \fboxsep0.75pt
    \caption{\textit{Number of Visual Examples.} Increasing the number of visual examples can introduce more variety for a more robust optimized condition at the expense of increasing optimization time, which can affect the performance negatively. \textbf{Best} results are bolded; the final setup is \colorbox{lightgray!20}{highlighted}.}
    \setlength\tabcolsep{8pt}
    \vspace{-3mm}
    \begin{tabularx}{\columnwidth}{X*{4}{c}}
        \toprule
         \multirow{2}{*}{Num Samples} & \multicolumn{2}{c}{SR} & \multicolumn{2}{c}{DeRain} \\
          {} & FID $\downarrow$ & LPIPS $\downarrow$ & FID $\downarrow$ & LPIPS $\downarrow$\\
        \midrule
        \rowcolor{lightgray!20}1 & \textbf{41.41} & 0.2666 & 87.12 & 0.2559\\
        4 & 46.24&0.2668 & 72.94&\textbf{0.2197} \\
        \rowcolor{lightgray!20}8 & 45.89&\textbf{0.2654} & \textbf{69.52}&0.2214  \\
        16 & 47.30&	0.2735 & 71.16 & 0.2227 \\
        
        \bottomrule
    \end{tabularx}
    \label{tab:exnum-ab}
\end{table}

\begin{table*}[t]
    \centering
    \footnotesize
    \fboxsep0.75pt
    \caption{\textit{Ablations on sampling steps and steering condition scale.}
    \textit{(a)} Increasing DDIM sampling steps improves editability but also introduces more optimization constraints. 
    \textit{(b)} Increasing the steering condition scale allows stronger deviations from the generative path, enhancing edit strength at the cost of fidelity. 
    \textbf{Best} results are bolded, and the final chosen configuration is \colorbox{lightgray!20}{highlighted}.}

    \setlength\tabcolsep{6pt}
    \vspace{-3mm}
    \begin{subtable}[t]{\columnwidth}
        \subcaption{\textit{DDIM Sampling Steps.}}
        \label{tab:ddimsteps-ab}
        \begin{tabularx}{\textwidth}{X*{4}{c}}
            \toprule
            \multirow{2}{*}{Steps} & \multicolumn{2}{c}{SR} & \multicolumn{2}{c}{DeRain} \\
            {} & FID $\downarrow$ & LPIPS $\downarrow$ & FID $\downarrow$ & LPIPS $\downarrow$\\
            \midrule
            50 & 45.43 & 0.2858 & 87.41 & 0.2566\\
            \rowcolor{lightgray!20}100 & \textbf{41.41} & \textbf{0.2666} & \textbf{87.12} & \textbf{0.2559} \\
            200 & 49.79 & 0.2815 & 91.94 & 0.2598 \\
            \bottomrule
        \end{tabularx}
    \end{subtable}
    \hfill
    \begin{subtable}[t]{\columnwidth}
        \subcaption{\textit{Steering Condition Scale.}}
        \label{tab:scale-ab}
        \begin{tabularx}{\textwidth}{X*{4}{c}}
            \toprule
            \multirow{2}{*}{Scale} & \multicolumn{2}{c}{SR} & \multicolumn{2}{c}{DeRain} \\
            {} & FID $\downarrow$ & LPIPS $\downarrow$ & FID $\downarrow$ & LPIPS $\downarrow$\\
            \midrule
            5 & 47.17 & 0.2801 & 89.24 & 0.2612\\
            \rowcolor{lightgray!20}7 & \textbf{41.41} & \textbf{0.2666} & \textbf{87.12} & \textbf{0.2559} \\
            9 & 45.73 & 0.2877 & 92.62 & 0.2733 \\
            \bottomrule
        \end{tabularx}
    \end{subtable}
\end{table*}

\begin{figure}
    \centering
    \includegraphics[width=\linewidth]{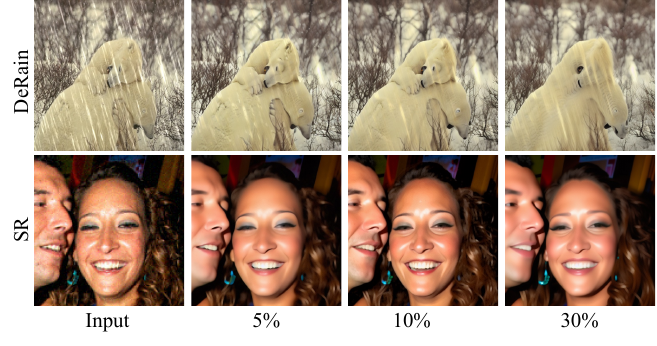}
    \vspace{-8mm}
    \caption{\textit{Diffusion path length effect}. Extending the diffusion path
increases variation, resulting in undesirable edits, while decreasing the path limits edibility. }
    \label{fig:str-ab}
\end{figure}

\begin{figure}[t]
    \centering
    \includegraphics[width=\columnwidth]{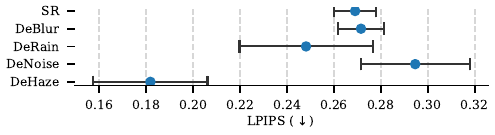}
    \vspace{-8mm}
    \caption{\textit{Sensitivity analysis.} We assess sensitivity by optimizing 10 models on unique examples; reporting the variance per task.}
    \label{fig:sensitivity}
\end{figure}

\paragraph{Performance Stability.}
\cref{fig:sensitivity} reports the variance across 10 models optimized on distinct reference pairs.
Despite minor fluctuations in complex tasks (DeRain), performance remains robust regardless of the chosen example.
This aligns with Fig.~\ref{fig:vis-conds}, which shows optimized conditions for the same task are closely similar regardless of the chosen visual example, proving reliable single-shot extraction.

\paragraph{Number of Visual Examples.} 
Increasing the number of visual examples helps optimize a more robust steering condition, especially for tasks with complex and highly variable patterns such as deraining. 
As shown in Tab.~\ref{tab:exnum-ab}, performance improves on the DeRain task as the number of examples increases. However, more examples also raise the optimization burden. When additional examples do not introduce new visual patterns, the added complexity can negatively impact performance.

\paragraph{DDIM Sampling Steps.} 
Using more DDIM sampling steps provides additional opportunities to apply edits, improving the method’s editability. However, increasing the sampling length also expands the number of conditions that must be optimized, making optimization more difficult and potentially degrading results. This trade-off is evident in Tab.~\ref{tab:ddimsteps-ab}.

\paragraph{Steering Condition Scale.} 
A higher steering condition scale increases the allowed deviation from the unconditioned generative trajectory (i.e., deviation from the input), which boosts editing strength at the cost of fidelity. As shown in Tab.~\ref{tab:scale-ab}, a small scale limits the model’s ability to apply the desired edits, while an excessively large scale expands the output space too aggressively, reducing performance.

\paragraph{Diffusion path length effect.} 
As discussed in Tab.~\ref{tab:stength-ab}, starting the sampling process deeper in the diffusion trajectory injects more noise into the latent, enlarging the output space but lowering fidelity. This effect is visible in Fig.~\ref{fig:str-ab}, where a longer diffusion path introduces unwanted content changes. Conversely, using too short a path overly restricts the output space, preventing the model from reaching suitable solutions and resulting in suboptimal edits.

\section{Limitations}
\label{sec:supp:limitations}

Our method leverages strong generative priors to handle complex edits on real images, but its performance ultimately depends on the capabilities of the underlying generative model. Although we move beyond the limitations of text-based conditioning to operate entirely in the visual domain, our results still reflect the strengths and weaknesses of this visual latent space. As seen in Figs.~\ref{fig:exp:sr}–\ref{fig:exp:color}, some fine textures may be lost due to the limited generative fidelity of Stable Diffusion~\cite{rombach2022high}, particularly when editing images processed through inversion. These limitations can be mitigated by adopting a more capable generative model, as demonstrated in Tab.~\ref{tab:bench-sana}.

However, latent diffusion models introduce an additional constraint: images are compressed into latent representations that may lose fine details. This affects both reconstruction quality and the ability to recognize subtle visual features. For instance, in Tab.~\ref{tab:bench-sana}, SANA-based methods underperform on the DeRain task due to SANA’s higher compression ratio. Employing a latent encoder specifically optimized for detail preservation could alleviate this issue.

Additionally, VDC prioritizes structural fidelity over non-rigid flexibility to prevent hallucinations, which limits large changes. Moreover, complex patterns (e.g., generalization to real rain) can challenge one-shot alignment. However, Tab.~\ref{tab:OOD-ab} confirms that simply adding more visual examples (synthetic) effectively mitigates this.

\section{Visual Results}
\label{sec:supp:visuals}

\begin{figure*}[t]
    \centering
    \includegraphics[width=\textwidth]{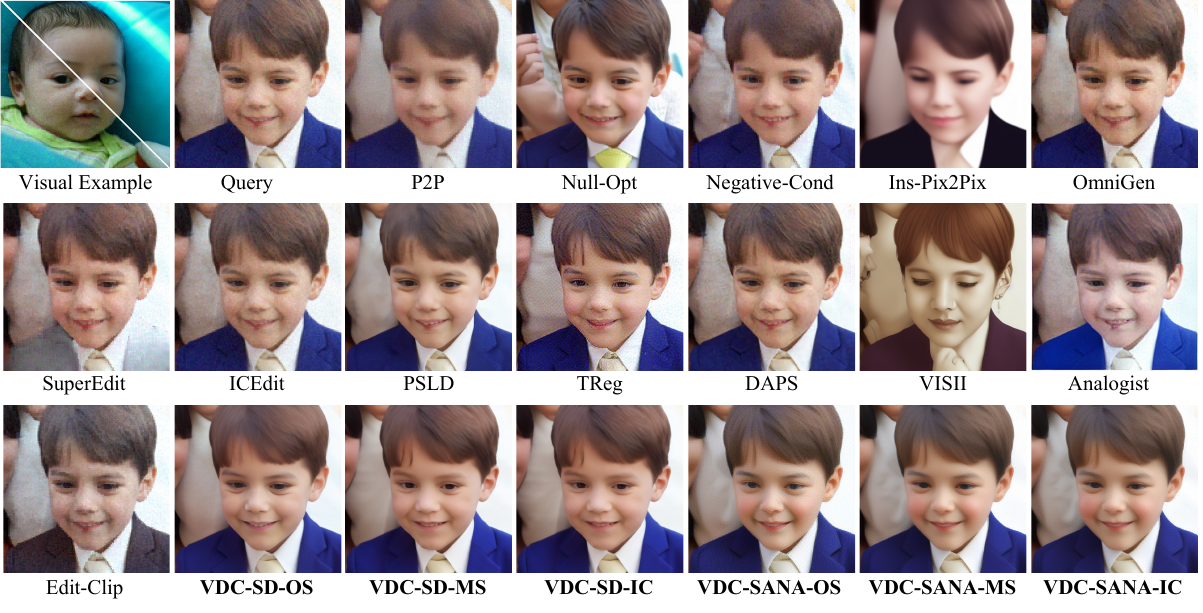}
    \vspace{-3mm}
    \caption{\textit{Visual comparison on SR task.} Text- and example-based approaches either fail to recognize the required edits or produce undesired changes and artifacts in the output. Our one-shot (OS) VDC yields clean results, with multi-shot (MS) and inversion correction (IC) modules improving generalization and fidelity.}
    \label{fig:exp:sr}
    \vspace{2mm}
\end{figure*}

\begin{figure*}[t]
    \centering
    \includegraphics[width=\textwidth]{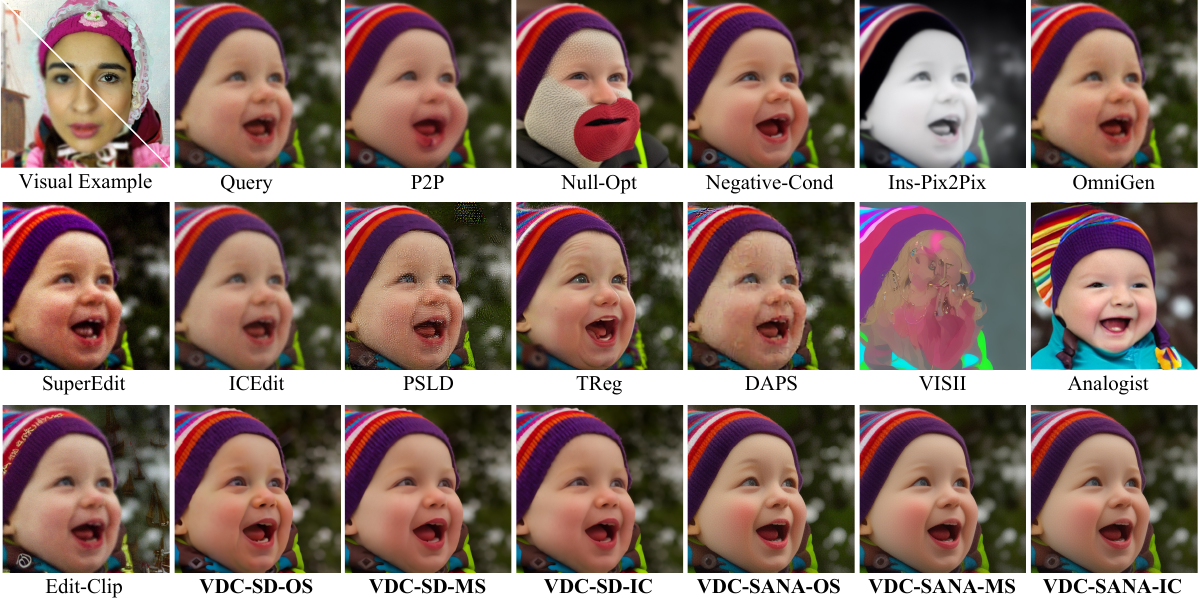}
    \vspace{-3mm}
    \caption{\textit{Visual comparison on DeBlurring task.} Text- and example-based approaches either fail to recognize the required edits or produce undesired changes and artifacts in the output. Our one-shot (OS) VDC yields clean results, with multi-shot (MS) and inversion correction (IC) modules improving generalization and fidelity.}
    \label{fig:exp:blur}
\end{figure*}

\begin{figure*}[h]
    \centering
    \includegraphics[width=\textwidth]{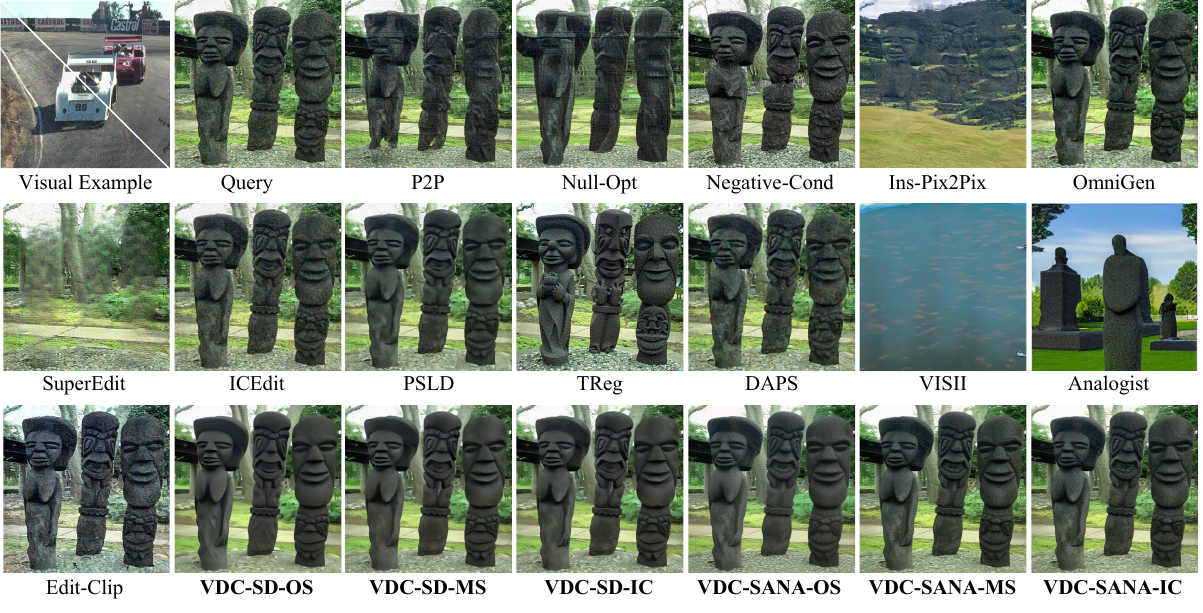}
    \vspace{-3mm}
    \caption{\textit{Visual comparison on DeNoising task.} Text- and example-based approaches either fail to recognize the required edits or produce undesired changes and artifacts in the output. Our one-shot (OS) VDC yields clean results, with multi-shot (MS) and inversion correction (IC) modules improving generalization and fidelity.}
    \label{fig:exp:noise}
    \vspace{2mm}
\end{figure*}

\begin{figure*}
    \centering
    \includegraphics[width=\textwidth]{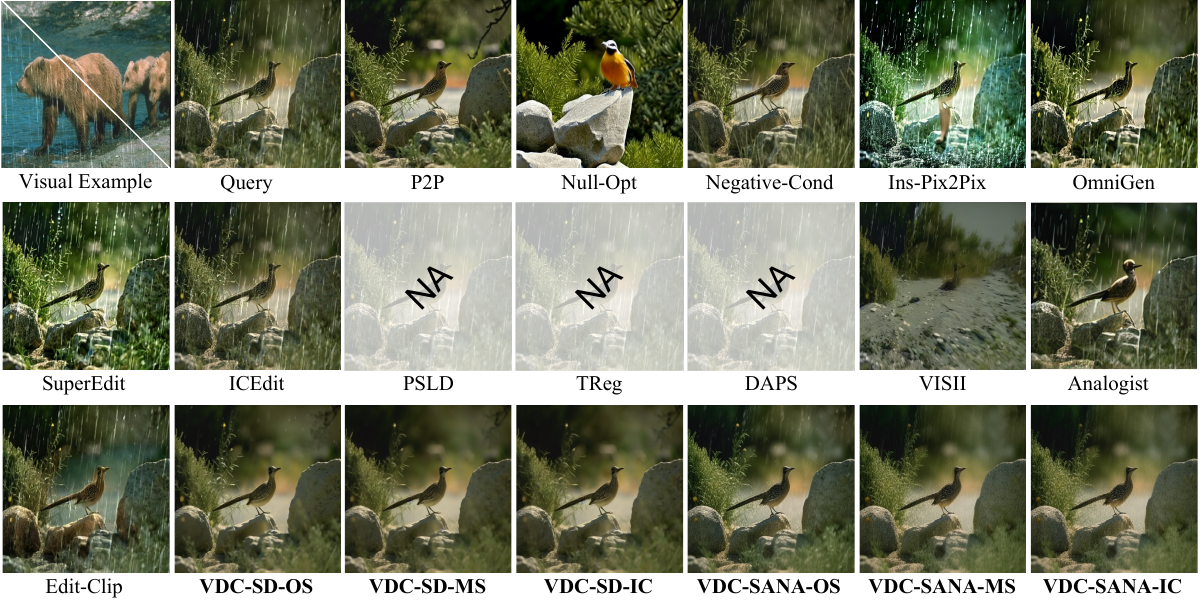}    
    \vspace{-3mm}
    \caption{\textit{Visual comparison on DeRaining task.} Text- and example-based approaches either fail to recognize the required edits or produce undesired changes and artifacts in the output. Our one-shot (OS) VDC yields clean results, with multi-shot (MS) and inversion correction (IC) modules improving generalization and fidelity.}
    \label{fig:exp:rain}

\end{figure*}

\begin{figure*}[h]
    \centering
    \includegraphics[width=\textwidth]{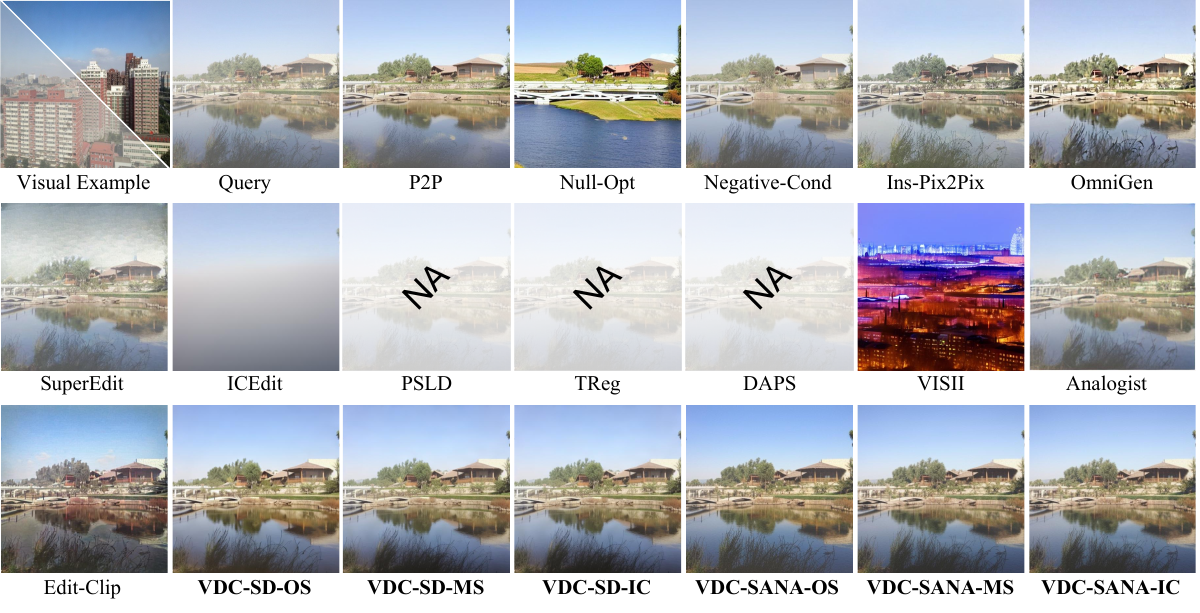}
    \vspace{-3mm}
    \caption{\textit{Visual comparison on DeHazing task.} Text- and example-based approaches either fail to recognize the required edits or produce undesired changes and artifacts in the output. Our one-shot (OS) VDC yields clean results, with multi-shot (MS) and inversion correction (IC) modules improving generalization and fidelity.}
    \label{fig:exp:fog}
    \vspace{2mm}
\end{figure*}

\begin{figure*}
    \centering
    \includegraphics[width=\textwidth]{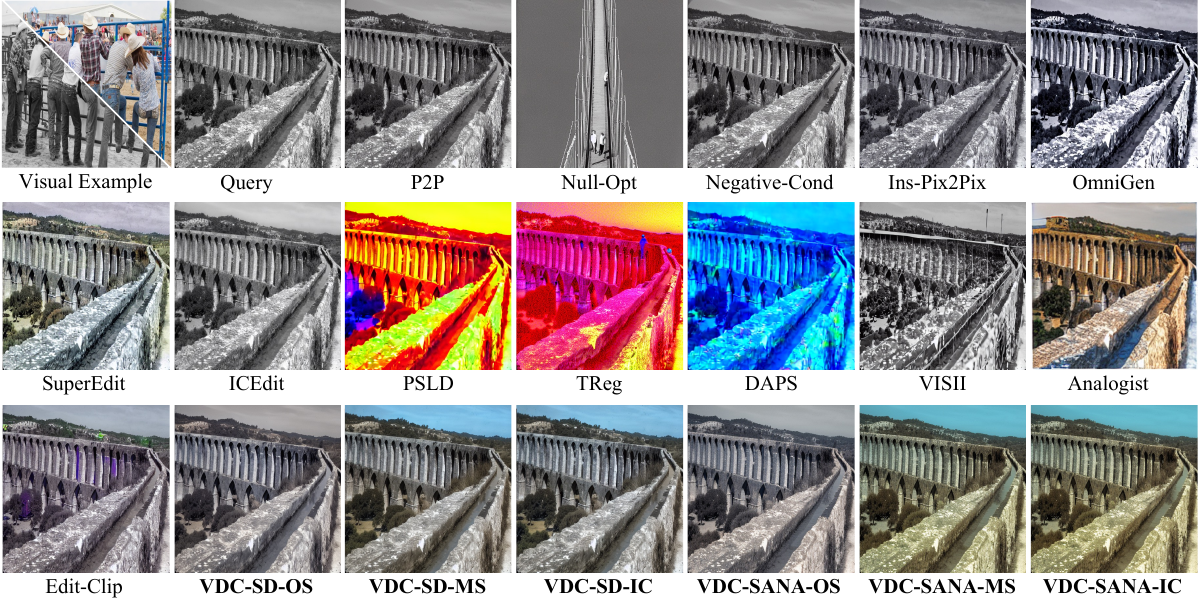}
    \vspace{-3mm}
    \caption{\textit{Visual comparison on Colorization task.} Text- and example-based approaches either fail to recognize the required edits or produce undesired changes and artifacts in the output. Our one-shot (OS) VDC yields clean results, with multi-shot (MS) and inversion correction (IC) modules improving generalization and fidelity.}
    \label{fig:exp:color}

\end{figure*}

In Figs.~\ref{fig:exp:sr}–\ref{fig:exp:color}, we provide additional visual comparisons across all baseline methods, as well as all variants of our approach using different generative models—Stable Diffusion (SD) and SANA—and different setups: One-Shot (OS), Multi-Shot (MS), and Multi-Shot with Inversion Correction (MS+IC).

\hfill

\hfill

\hfill
